\documentclass[useAMS,usenatbib]{mn2e}
\usepackage{epsfig}
\usepackage{color}
\usepackage{float}
\usepackage{times}
\usepackage{longtable}
\usepackage{float}
\usepackage{graphicx} 
\usepackage{epstopdf}
\usepackage{amsmath}
\usepackage{threeparttable }
\usepackage{lineno}
\usepackage{hyperref}
\hypersetup{colorlinks=true,linkcolor=blue,citecolor=blue,filecolor=blue,urlcolor=blue}

\begin{document}

\title[Radio transients and variables classification] 
{On the use of variability time-scales as an early classifier of radio transients and variables}
 \author[Pietka, Staley, Pretorius \& Fender]
       {M. Pietka$^{1,2}$\thanks{email:malgorzata.pietka@physics.ox.ac.uk}, T. D. Staley$^{1,2}$, M. L. Pretorius$^{1,2,3}$, R. P. Fender$^{1,2}$\\ 
$^1$ Astrophysics, Department of Physics, University of Oxford, Keble Road, Oxford OX1 3RH, UK\\
$^2$ Physics and Astronomy, University of Southampton, Highfield, Southampton, SO17 1BJ, UK\\
$^3$ South African Astronomical Observatory, PO Box 9, 7935 Observatory, Cape Town, South Africa\\ }
\maketitle

\begin{abstract}
We have shown previously that a broad correlation between the peak radio 
luminosity and the variability time-scales, approximately $L \propto \tau^{5}$,
exists for variable synchrotron emitting sources and that different classes of astrophysical source occupy different regions of luminosity and time-scale space.
Based on those results, we investigate whether the most basic information available for a newly discovered radio variable or transient -- their rise
and/or decline rate -- can be used to set initial constraints on the class of events from which they originate.
We have analysed a sample of $\approx$ 800 synchrotron flares, selected from light-curves of $\approx$~90 sources observed at 5-8 GHz, representing a wide range of astrophysical phenomena, from flare stars to supermassive black holes. Selection of outbursts from the noisy radio light-curves has been done automatically in order to ensure reproducibility of results.  The distribution of rise/decline rates for the selected flares is modelled as a Gaussian probability distribution for each class of object, and further convolved with estimated areal density of that class in order to correct for the strong bias in our sample.
We show in this way that comparing the measured variability time-scale of a radio transient/variable of unknown origin can provide an early, albeit approximate, classification of the object, and could form part of a suite of measurements used to provide early categorisation of such events. Finally, we also discuss the effect scintillating sources will have on our ability to classify events based on their variability time-scales.
\end{abstract}

\begin{keywords} 
radiation mechanisms: non-thermal; radio continuum: transients; methods: observational
\end{keywords}

\section{Introduction}
\label{intro}

In the coming years, the next generation of radio telescopes are expected to survey large areas of sky to a sensitivity which could see the discovery of hundreds of transient and variable sources  \citep{Fender2015, Metzger2015}. 
Previous discoveries of radio transients proved that relying solely on radio data makes it extremely difficult to validate both the nature, as well as the reality of the source (\citealt{Bower2007}; \citealt{Stewart2016}) and that it is vital they are supported by simultaneous multiwavelength observations or rapid follow up.  
On the other hand, with a rate of transient/variable candidates as high as expected, it will not be possible to follow up each event with other multiwavelength instruments. However, if the nature of the source is constrained as early as possible, decision-making strategies can be created to follow up the type of events relevant to specific science goals.

The importance of the automatic classification of light-curves has been discussed before in the context of deep surveys carried at optical wavelengths, with attempts to include automatic classification of transients into optical surveys (\citealt{saglia2012, djorgovski2012}). Extended investigations of different ways to approach this task include developing classification techniques based on machine learning methods \citep{richards}, Bayesian theory \citep{pichara}, density approach (eliminating direct measurements of features such as magnitude or color; \citealt{kugler}), fitting non-complete light-curves \citep{Lo2014} and other non-parametric techniques \citep{varughese}. 
Those methods have been developed using archival optical light-curves as training datasets, as well as X-ray transients (\citealt{Lo}, with a supervised learning technique implemented). For radio data, \citet{rebbapragada} test different classification algorithms on simulated light-curves in preparation for the ASKAP VAST survey. Most of these methods, although they work effectively on sparse and noisy datasets, require the light-curve/flare to be (mostly) complete. In this work, however, we are focused on investigating whether an initial classification of the source based on the partial information about its light-curve is possible. 

For any transient or variable candidate discovered in a blind radio survey, the most basic measurable property of the light-curve is its variability time-scale. 
Here we present a potential technique of classifying these types of events by measuring the rise/decline rates of their flares, which could be used as the early step in a more complex classification pipeline. At the moment however, due to the relatively small data sample used to develop the method, it is not a final classification solution.

As we have previously shown (\citealt{paper1}; hereafter PFK15), there is a correlation between the luminosity ($L$) and rise/decline rates ($\tau$) that exists between different classes of synchrotron emitting sources, approximately  $L \propto \tau^{5}$. 
It covers a wide range of events, from nearby flare stars to extragalactic supermassive black holes, with the more massive and luminous objects displaying variability on progressively longer time-scales. 
Although it has been expected previously that more massive and luminous objects vary on much longer time-scales compared to intrinsically faint sources \citep{vdl}, the analysis presented in PFK15 showed a clear distinction between different classes of objects occupying the luminosity -- time-scale parameter space. 
Because at the time of discovery of a radio event its distance is unknown (unless a counterpart at other wavelengths is immediately identified), the information about the luminosity is not available, and the discussed relation (Figure~3 in PFK15) is reduced to a distribution of the variability time-scales.
This result offers an opportunity to develop a method which could be used to perform an initial classification of an unknown object based on its radio data only. 
Approximate classification by time-scale and luminosity has its analogues in the field of optical transients (e.g. Fig 1 of \citealt{Rau2009}).

It should be noted, that the results presented in PFK15 were based on the analysis of a sample of single flaring events, manually selected from various radio light-curves. However, if we were to base an early classification on those results, the analysis should be done in a reproducible way, such that the bias introduced by identifying flares by eye is minimized. In order to do that, we have developed a basic piece of software which automatically identifies and selects flaring events from the radio light-curves in our sample. 
Our routine is based on a simple thresholding approach, where any variability of the light-curve above a chosen flux density threshold is defined as a `flare'.  However, this routine is designed to work specifically on the diverse dataset analysed in this study.  
In future surveys, where data quality is more consistent, complex algorithms will be more appropriate to be used as part of the detection pipeline.

An important aspect of the presented analysis which needs to be carefully addressed is the bias associated with our sample of radio light-curves. The number of objects analysed within each class represents the frequency and quality of observations rather than the actual sky density of those sources. 
Therefore, in order to accurately predict the probability of finding a given class of objects, this effect needs to be accounted for. Although the exact areal densities of analysed classes of objects are not well known, we attempt to estimate those values and convolve them with the obtained distribution of rise/decline rates across a range of considered classes. This final distribution provides an initial reference point against which we can compare the measurements of time-scales taken for any newly discovered transient/variable candidate.

In Section~\ref{data} we describe the sample of light-curves used in the analysis. Section~\ref{flarefinding} provides details of the simple flare finding routine we developed to select flares from our sample of light-curves, as well as an overview of alternative methods for the flare selection. We describe the analysis of selected flares in Section~\ref{analysis} and give details of the estimation of expected areal densities in Section~\ref{se}. Final results are presented in Section~\ref{results} and discussed in Section~\ref{discussion}. Conclusions are given in Section~\ref{conclusions}.

\section{Data}
\label{data}
The data included in the analysis span a broad range of radio variable sources, 
observed at frequencies between 5 and 8 GHz, compiled from the literature and the Green Bank Interferometer (GBI) archive\footnote{\url{ftp://ftp.gb.nrao.edu/pub/fghigo/gbidata/gdata/}}.
Overall, the analysed light-curves originate from a range of instruments and different observing programmes, therefore, there is no consistency in their sampling intervals, length of time series or sensitivities. 
The number of sources from different classes is as follows: AGN (28), tidal disruption events (2), supernovae (13), GRB afterglows (4), classical novae (7), dwarf nova (1), X-ray binaries (18), magnetar (1), RSCVn (3), algol (3), magnetic CVs (2) and flare stars (4), giving the total number of light-curves from synchrotron emitting sources of 92.
A detailed list of sources included in the sample, together with corresponding references, can be found in PFK15, with the exception of one tidal disruption event not included in the previous work - details of this source are summarized in Table~\ref{tdesample}.
Additionally, listed in Tables~\ref{sctsample} and \ref{esesample} are scintillating sources and extreme scattering events (ESE). Because the discussed luminosity--time-scale correlation applies to intrinsic variability only, these sources are not included in the main analysis. However, we use this sample to investigate the time-scales of extrinsic variability and discuss the extent of overlap with time-scales of synchrotron flares (Section~\ref{discussion}).
The errors of the flux density measurements have been provided for all the GBI data sets, and listed in 
selected papers. In cases where the errors were not specifically stated, we have estimated 
them based on the published figures, which, in majority of the cases meant taking ten per cent of the flux density measurement. 

The following section describes the basic flare finding software, which automatically selects flaring events 
from the described sample of radio light-curves, in order to analyse variability time-scales across the classes of objects.

\section{Automatic selection of flares}
\label{flarefinding}

\subsection{Overview of flare identification methods}

There are a number of possible approaches to measuring time-scales, and in particular flaring time-scales, in time-series data. Perhaps the simplest and most intuitive approach is to adopt a given threshold and designate any contiguous sections of the time-series above that threshold as a `flare'. A time-scale measurement can then be obtained by measuring the time spent above the threshold level, but of course this requires that we wait for the flare to subside. Alternatively we may attempt to fit a model of some kind to any data points already recorded. This approach has often been adopted in the past due to its relative simplicity, robustness, and ease of implementation. Depending on the model-fitting requirements the thresholding approach can also be extremely computationally efficient, and hence suitable for real-time applications. \\
However, there are several draw-backs to the basic thresholding approach. First and foremost, finding a suitable threshold level can be a somewhat arbitrary process. If sources of noise are well characterised then a threshold can be chosen on the basis of a calculated false-positive rate, but this may not be possible in all cases. Even if instrument characteristics are well known, an astronomical source may display low-level intrinsic variability which is qualitatively different to the rapid flux-rise-and-decay characteristic of a flaring event. As such it may be necessary to devise some kind of calibration technique (as we do in this work) or even to manually pick a suitable threshold level. The second problem is that of choosing a suitable model for flare-fitting, which again is often somewhat arbitrary, although hopefully motivated by knowledge of the physical processes at work. Additionally, simple thresholding provides no means of separating multiple-superimposed flares, which degrades the accuracy of any model fits. Finally, a point-wise threshold will always lose some sensitivity compared to more sophisticated approaches, if enough is known about the flare-morphology to apply an effective matched-filter. In the more general context of time-scale measurement, model-free metrics such as autocorrelation time and other metrics for aperiodic variability \citep{findeisen2015} can be applied, but typically require a large amount of variable time-series data to work effectively. Since we are interested in time-scales as an early-time classification tool, we do not consider these approaches further here.\\
More advanced change-or-flare-detection techniques typically take a probabilistic approach. As such they give a more rigorous and informative measure of the time-series data, at the cost of additional computational time and complexity. One possibility might be application of the Bayesian Blocks algorithm \citep{scargle1998} for a model-free probabilistic method of change-detection. This side-steps the problem of threshold-determination, but again does not provide an early-time estimate of characteristic time-scale. If a reasonable model can be chosen, then Monte-Carlo methods can be applied to estimate the likelihood of a flare presence with excellent sensitivity \citep{pitkin2014}, or even to dissect the time-series into multiple superimposed flares \citep{Huppenkothen2015}, though this requires significant computational time. A thorough investigation of the relative accuracy and efficacy of such methods would be interesting, but is outside the scope of this work. \\
In our case, advanced flare finding techniques are not suitable to work on the diverse sample of light-curves we analyse. For this study we apply a thresholding approach with a simple exponential model. This provides a reasonably robust time-scale estimation for modest computational effort, albeit with some limitations to applicability, as detailed below.

\begin{figure*}
\includegraphics[width=0.49\textwidth]{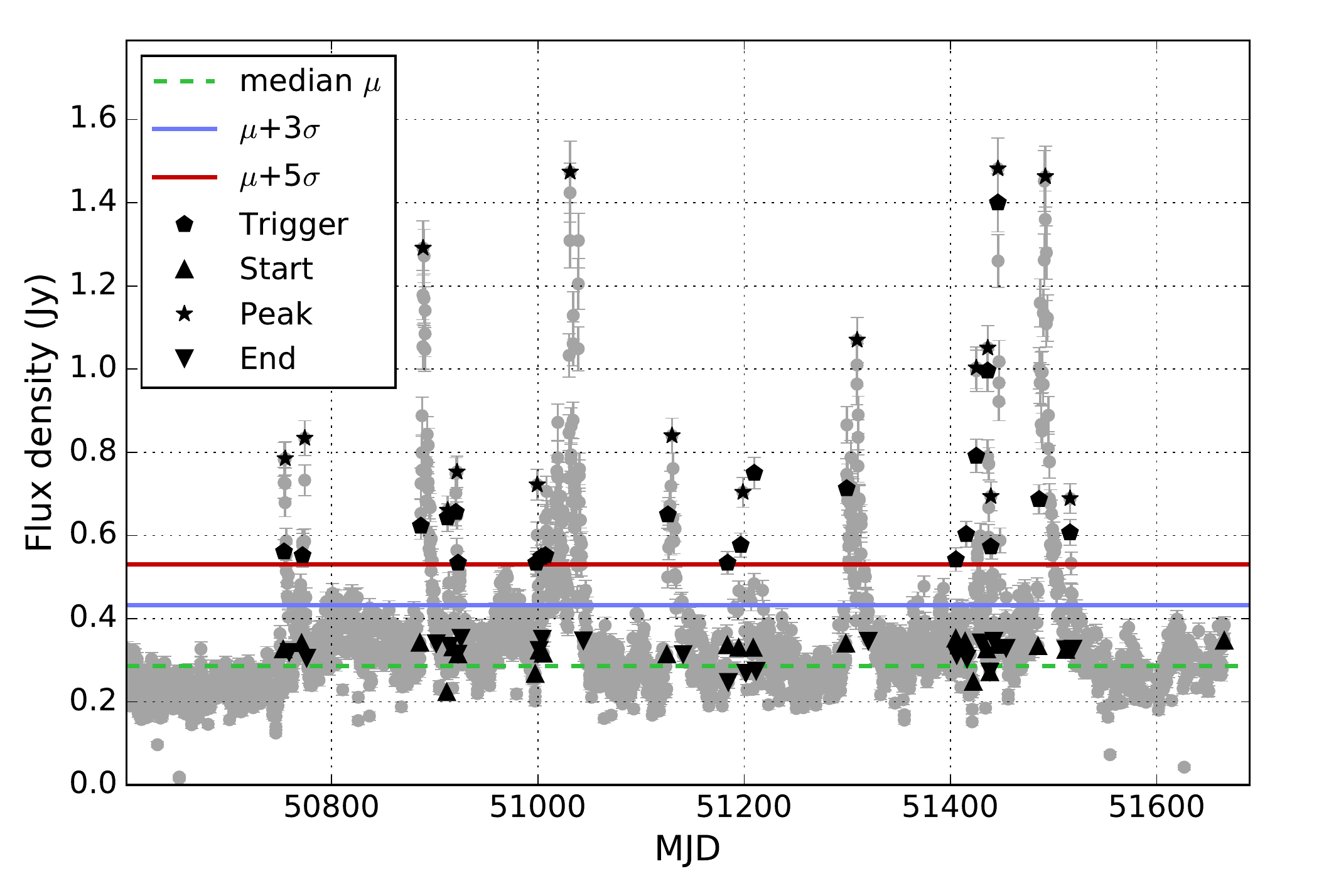}
\includegraphics[width=0.49\textwidth]{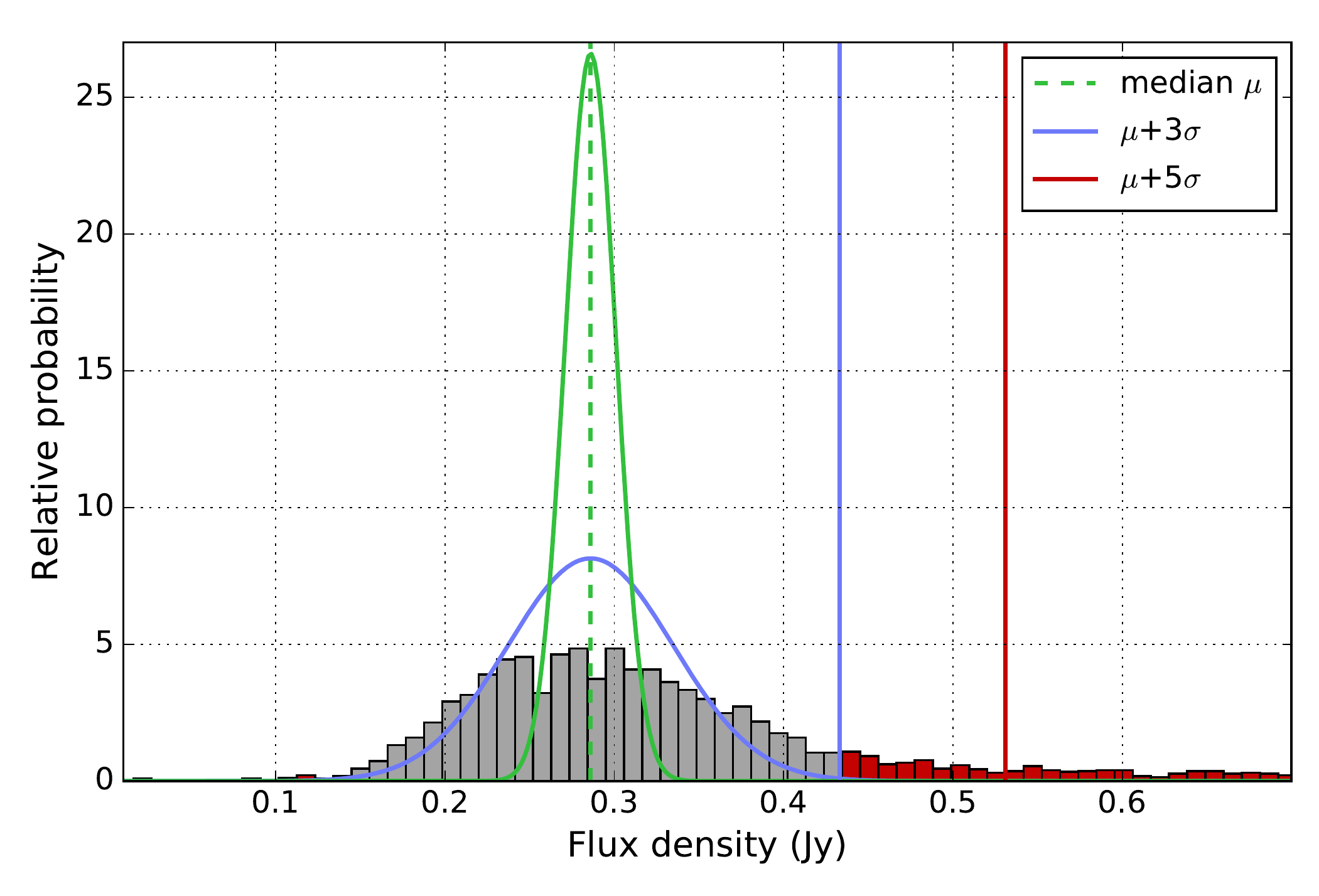}
\includegraphics[width=\textwidth]{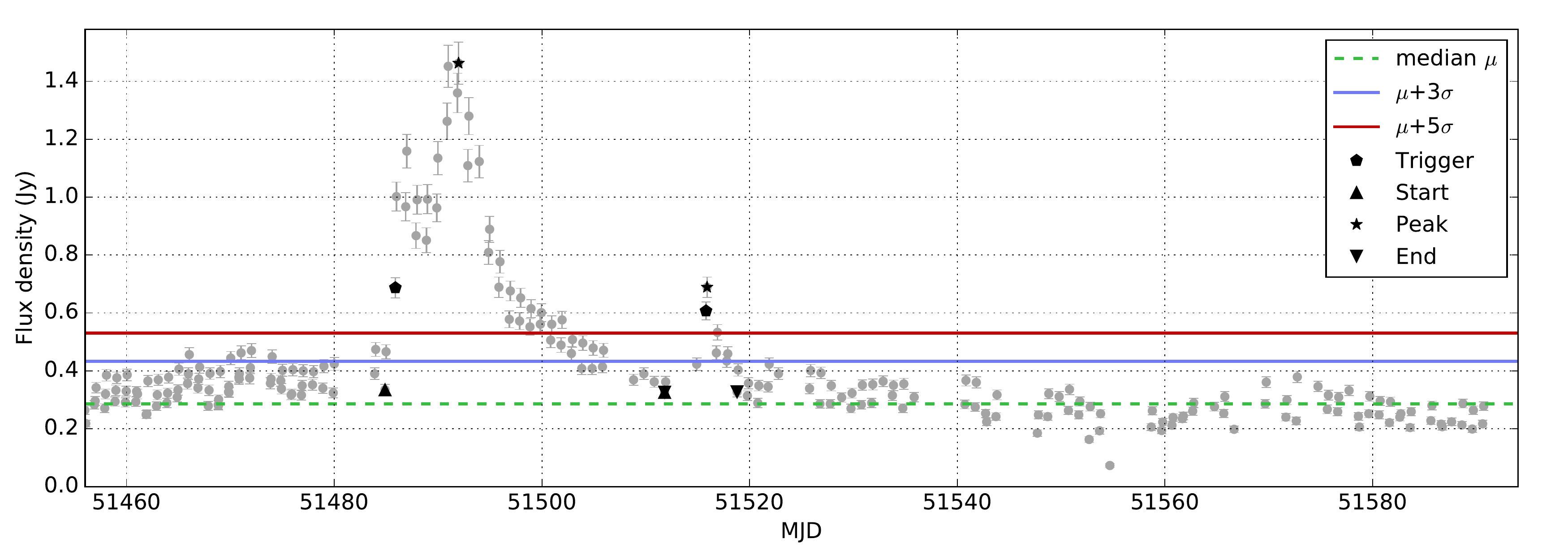}
\caption{Example figure showing the detection of flaring events in the radio light-curve of the binary SS~433~(GBI). {\em Upper panel, left}: light-curve of the source and flaring events selected by the Automatic Flare Finder (AFF). Marked with the dashed green line is the estimated quiescent flux level (labelled `median' in the key). The blue and red lines indicate the variation above the quiescent level by 3$\sigma$ and 5$\sigma$ respectively. The black symbols on each flare mark: start/end of the flare (triangles), data point triggering the flare detection (pentagons) and the peak of the flare (stars). 
{\em Upper panel, right}: histogram of flux density values for the light-curve after sigma clipping has been applied. Flare-threshold values of 3$\sigma$ and 5$\sigma$ are marked by the blue and red vertical lines respectively. Curve overplotted in blue represents a Gaussian distribution with median and median absolute deviation matched to the sigma-clipped data. Curve overplotted in green represents a Gaussian distribution of the same median, but with quiescent variability matched to the formal error bars on the flux-density measurements, clearly showing that the formal error bars underestimate the low-level variability in this light-curve.
{\em Lower panel}: close up view of two flares selected from the light-curve, showing clearly the trigger, beginning, peak and end of the flare markers.}
\label{AFF}
\end{figure*}

\subsection{Automatic Flare Finder (AFF) for PFK15 dataset}
\label{aff}

The flare-identification routine we tested is simple, but nonetheless provided a reproducible means of characterising flare time-scales for this work\footnote{The flare finding software along with the subsequent scripts used to model the probability distributions is available at {\url{https://github.com/4pisky/variability-timescale-analysis-paper}}}.
Using more advanced methods discussed in the previous section would be impractical, since our datasets are very diverse (Section \ref{data}) and do not meet the requirements of such complex techniques.
It should be stressed that, while this simple method is sufficient to select flares from our sample of light-curves, it is not intended to be used as a generic tool.
It uses a simple thresholding approach, as follows:
given a pre-determined quiescent or background flux level estimate, $b$,
and an estimate of the signal variation due to noise, $\sigma$,
the routine first steps through the data looking for data points $x$ where the flux $f_x$ is more than $5\sigma$ above the quiescent flux:
\begin{equation}
x: \quad f_x > b + 5\sigma\, \textrm{;}
 \label{eq:threshold}
\end{equation}
these are referred to as the `trigger' points.
Once a trigger is found, the algorithm searches for the nearest data points before and after the trigger which have flux less than
$1\sigma$ above the quiescent flux-level, these are designated as the flare start and end.

The tricky part of this process is determining a suitable quiescent / background flux level, and estimating the levels of signal variation present when the source is not undergoing an outburst.
For a source which has a well-sampled long-term light-curve encompassing extended periods of quiescence, the first problem -- `background' flux-level estimation -- is straightforward; simply taking the median flux level provides a reasonable estimate.
However, when analysing a light-curve which displays flares for a significant fraction of data-period, the median may overestimate the background level.
The second problem, estimating the quiescent low-level variance, is harder.
Again considering the ideal case of a well-sampled long-baseline light-curve, at some level we could simply use the formal errors on the data points and designate anything greater than $5\sigma$ above the median a flare.
However, many of the light-curves analysed for this paper display low-level variation which does not subjectively qualify as a flare, but which is nonetheless larger than would be expected from the formal errors, and appears persistent over multiple data points (e.g. SS 433, as shown in Figure~\ref{AFF}) -- we designate this as intrinsic quiescent variation (though varying telescope systematic noise-levels cannot be ruled out without access to the original raw data).
To avoid the need to manually change the $\sigma$-threshold for each dataset according to the level of quiescent variation, we required an alternative method of estimating the low-level variation.
For this purpose, we employed the sigma-clipping routine from the {\em astropy} library \citep{astropy}, using the default clipping-threshold of $3\sigma$, and iterating until convergence.  This means estimating the median value of the whole light-curve, clipping all the points which are above $3\sigma$ and repeating the procedure until there are no points to reject. We have used median absolute deviation as a measure of the data variability $\sigma$.
For some of the light-curves we analysed this does a good job of masking the high-flux outlier data points representing flares, however, alternative methods of finding the low-level variation for light-curves which require more specific approach are discussed in Section~\ref{prep}.
We then estimate the median and median absolute deviation of the remaining unmasked `quiescent' data, and use these quantities for the values of $b$ and $\sigma$ in Equation~\ref{eq:threshold}. Figure~\ref{AFF} shows an example of the flares identified in a GBI light-curve for the X-ray binary SS~433.

\section{Analysis}
\label{analysis}

The quality of the data described in Section \ref{data} varies significantly across the sample.
Data provided by the GBI consists of noisy light-curves for both slowly (AGN) and rapidly 
varying sources (XRBs, RSCVn), observed on time-scale of years. Data compiled 
from the literature, although much less noisy, in many cases shows pre-selected flares, 
sometimes with the background emission subtracted (GRBs, SNe).

The diversity of available light-curves presented a substantial challenge in designing 
software in a way that would allow us to detect flaring events with no manual intervention 
during the process. Even the simple thresholding method described in Section \ref{aff} sets a number of constraints on the light-curve which make it difficult to implement across the variety of the datasets. 
Firstly, it requires the light-curve to be long enough in relation to the duration of the flare, 
such that the background flux level (excluding flares) can be accurately estimated. 
Secondly, it needs to be sufficiently noise-free in order to avoid false detections. In this section we describe the general 
	adjustments that have been made to the flare identification routine and/or data sample 
in order to make the selection of flares as optimal as possible.


\subsection{Data preparation}
\label{prep}
Testing the AFF on all the available light-curves identified a set of issues which allowed us to divide the whole data sample into four groups. Those groups, together with list of adjustments made and example plots are discussed below.

\subsubsection{Short duration flares selected from GBI (15\% of all light-curves; 84\% of analysed flares)}
\begin{figure}
\includegraphics[scale=0.415]{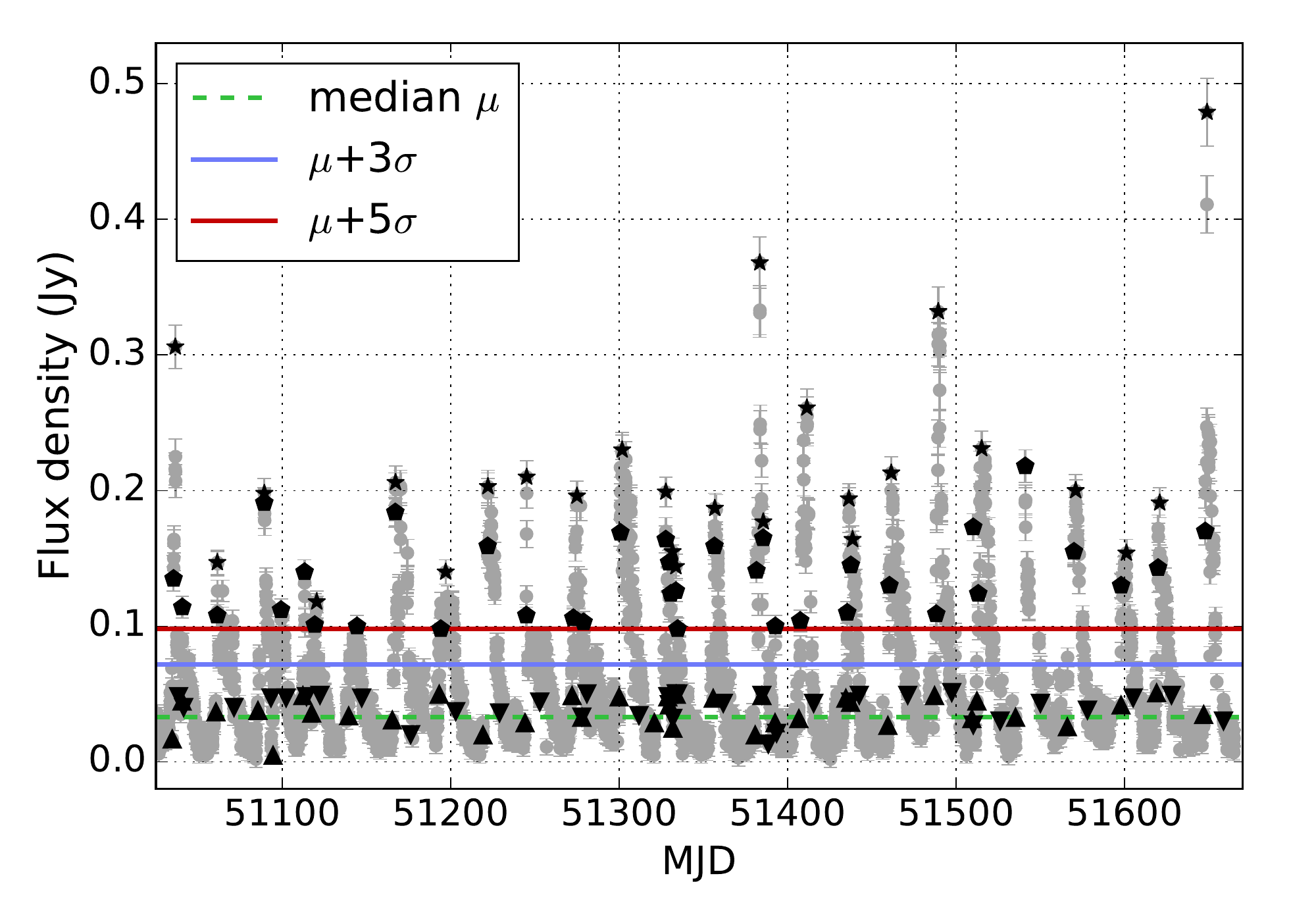}
\caption{An example of a light-curve showing multiple short flares for a periodic X-ray binary LS I +63$^{\circ}$303.
Flaring events selected from this 8.3~GHz GBI light-curve by the AFF are marked with the black stars. The background level has been estimated using sigma clipping method described in Section~\ref{aff}.}
\label{gbim}
\end{figure}

Light-curves in this group consist of multiple, short duration 
($\approx$~days) flares, observed on a time-scale of years.
This set of data as the only one in our sample presented no issues for the AFF. 
Figure~\ref{gbim} shows an example of such light-curve, with a number of 
identified flares.

\begin{figure}
\includegraphics[scale=0.415]{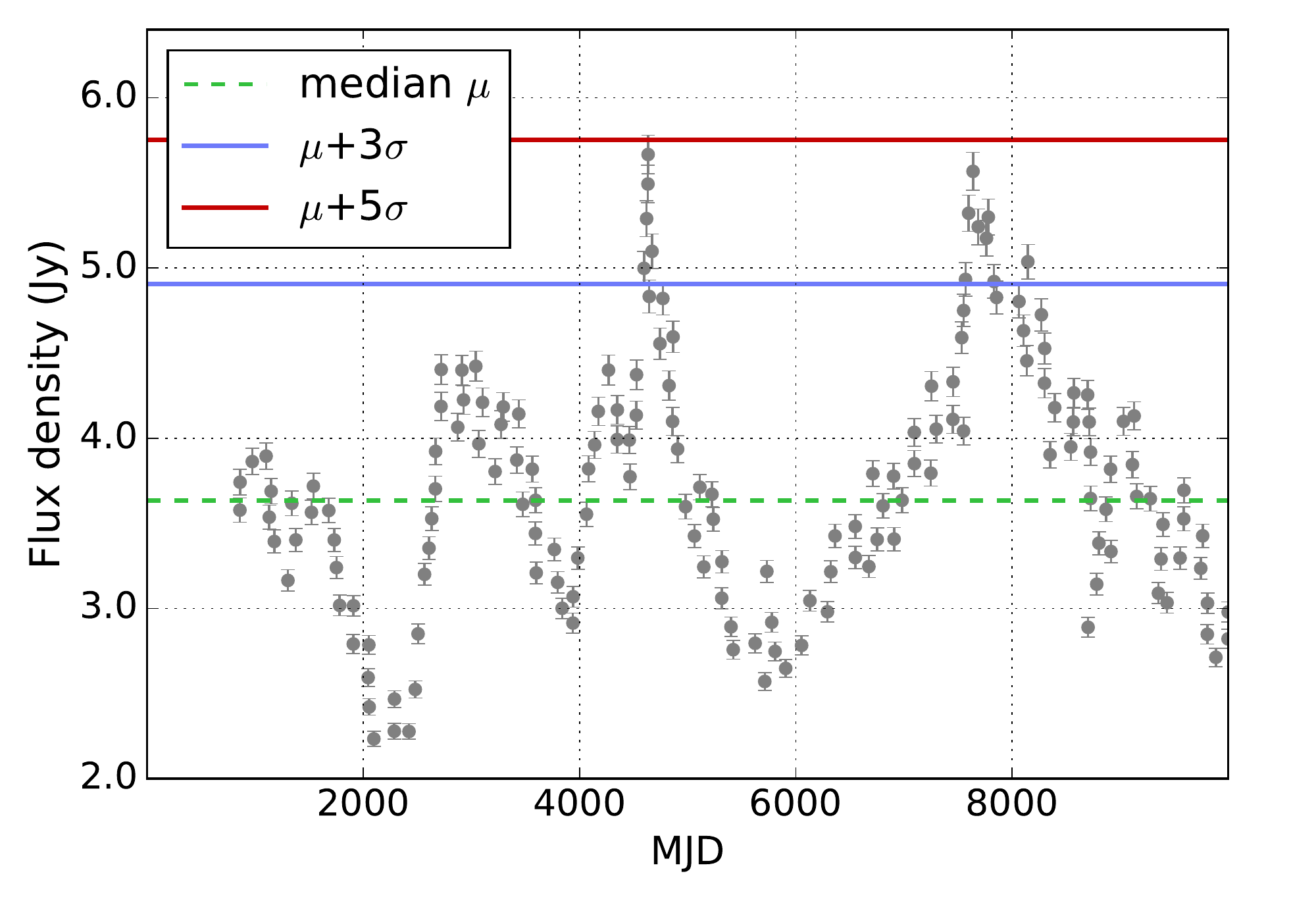}
\includegraphics[scale=0.415]{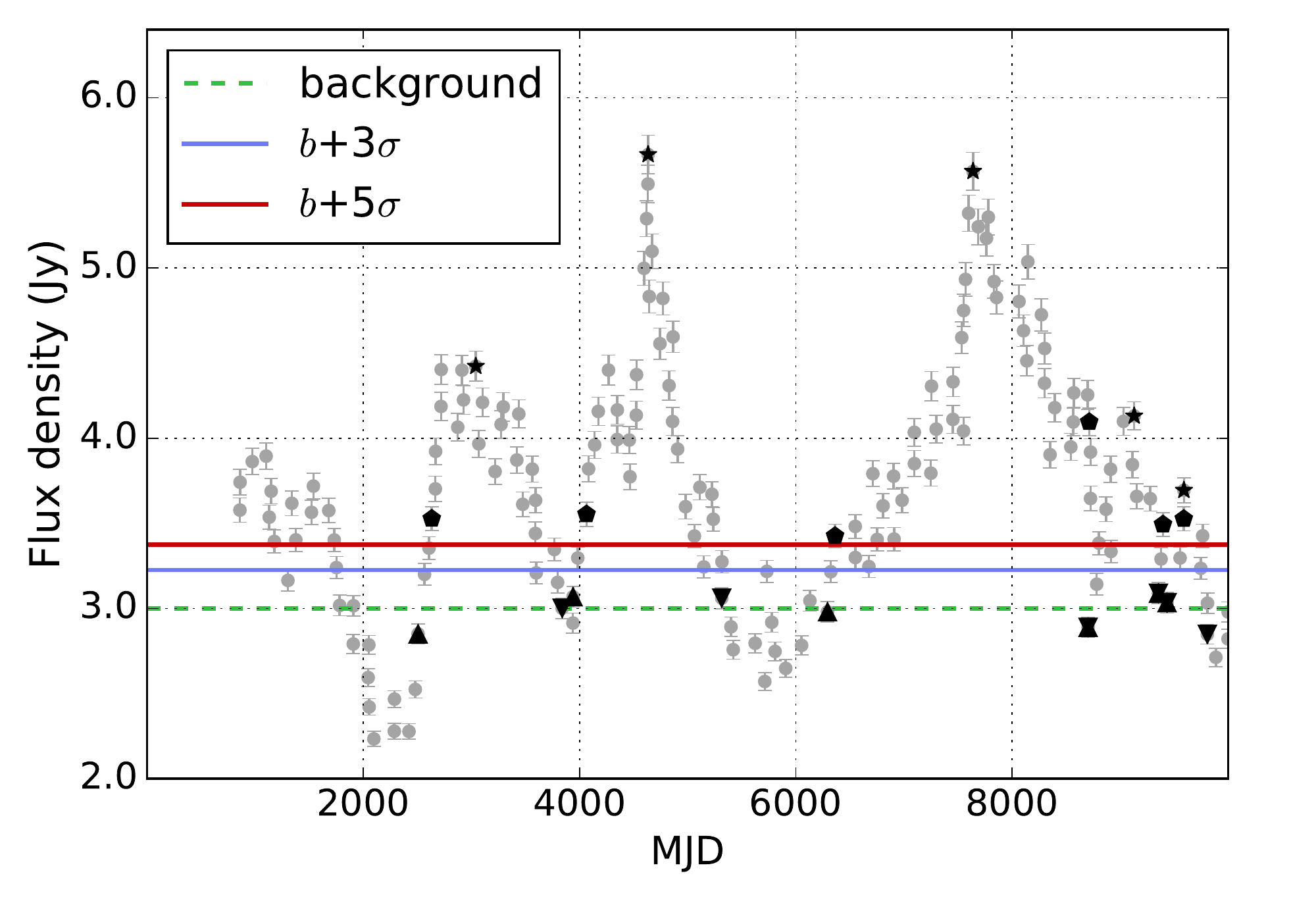}
\caption{{\em Upper}: An example of a light-curve showing multiple flares with the quiescent level insufficiently well characterised for the automatic software to identify flares. (3C120, \citealt{3C}). {\em Lower}: The same light-curve, where the quiescent flux level
has been estimated as the $15^{\mathrm{th}}$ percentile of the flux measurements, and, for calculating the flare-detection-trigger amplitude, we set $\sigma$ equal to the mean of the formal error bars on the flux density measurements.}
\label{litm}
\end{figure}

\begin{figure}
\includegraphics[scale=0.415]{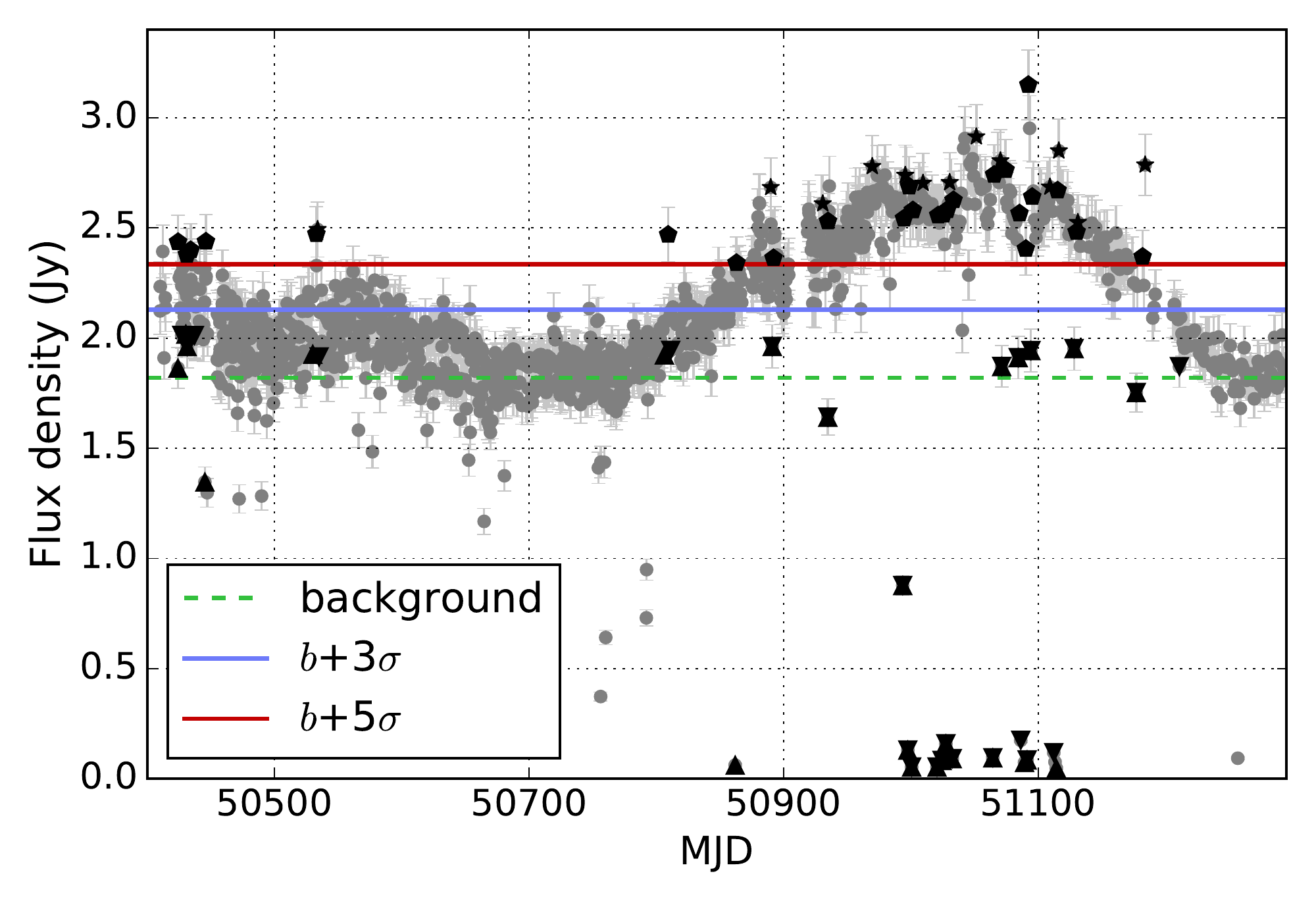}
\includegraphics[scale=0.415]{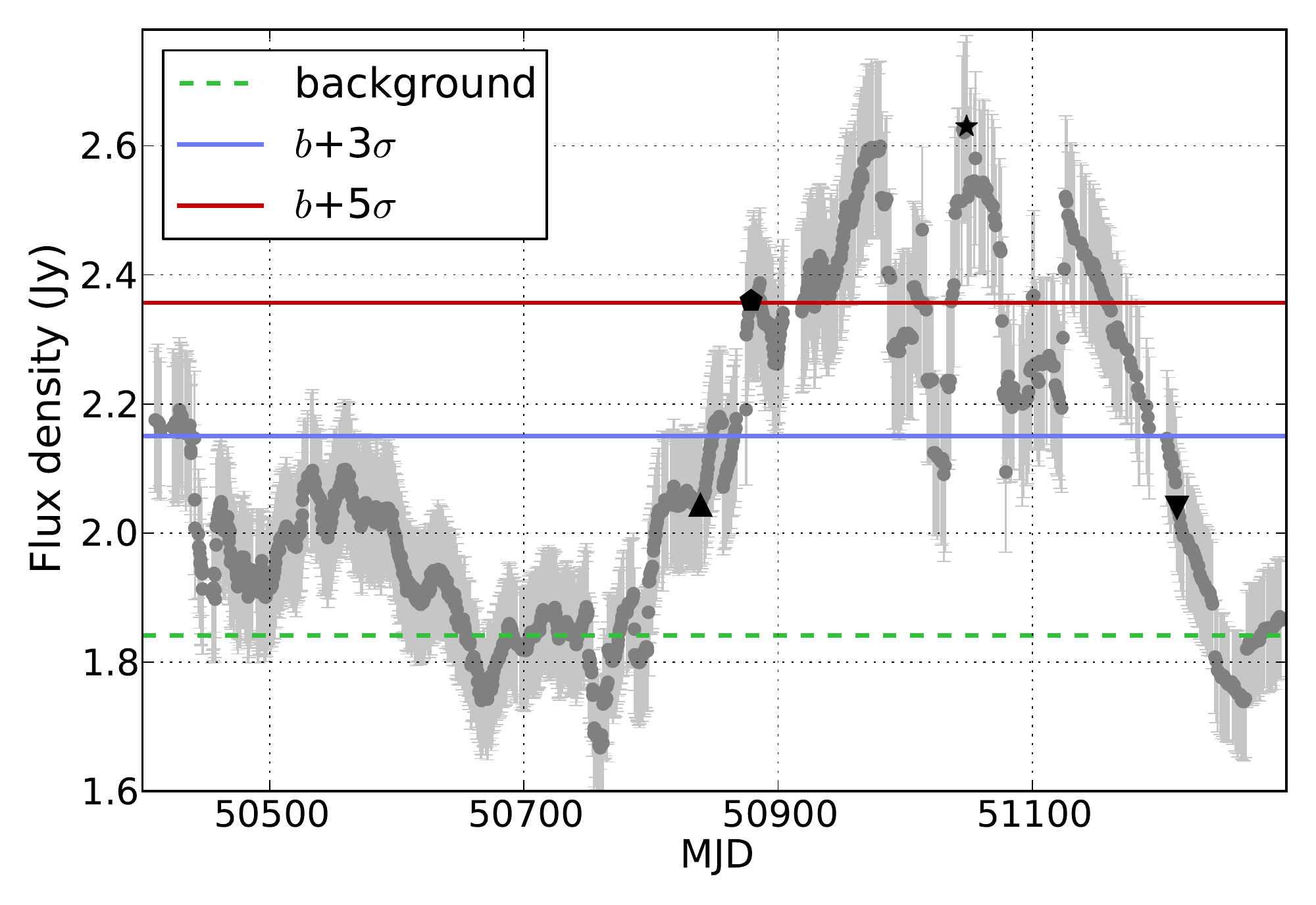}
\caption{{\em Upper}: An example of a light-curve showing long time-scale flares (0224+671, GBI). 
With the background and the quiescent low-level variability estimated as in lower panel of Figure~\ref{litm}, some odd noisy data points are falsely identified as flares. 
{\em Lower}: The same light-curve smoothed, with one major flare recognized by the software.}
\label{gbis}
\end{figure}

\subsubsection{Flares compiled from the literature (30\% of all light-curves; 9\% of analysed flares)}
\label{datalits}
This group comprises the light-curves of repeating, as well as several cataclysmic events.
However, information about the background level in this sample is limited to the close vicinity of the flares, 
and therefore is insufficient for the correct estimation of the background emission and the low-level variability.
In most cases, the automatically estimated values were too high when compared to the 
peak flux of the outbursts, and as a result, the flares were not recognized 
as varying above chosen (5$\sigma$) threshold. 
In order to resolve that problem, we have estimated the quiescent flux level of the light-curve as the
15$^{\mathrm{th}}$ percentile of the flux density measurements.
For calculating the flare-detection-trigger amplitude, we set $\sigma$ equal to the mean of the formal error bars on the flux density measurements. 
An example of such a light-curve, showing both original attempt to select flares by AFF, 
as well as flares selected with the alternative approach is shown in Figure~\ref{litm}.

\begin{figure}
\includegraphics[scale=0.415]{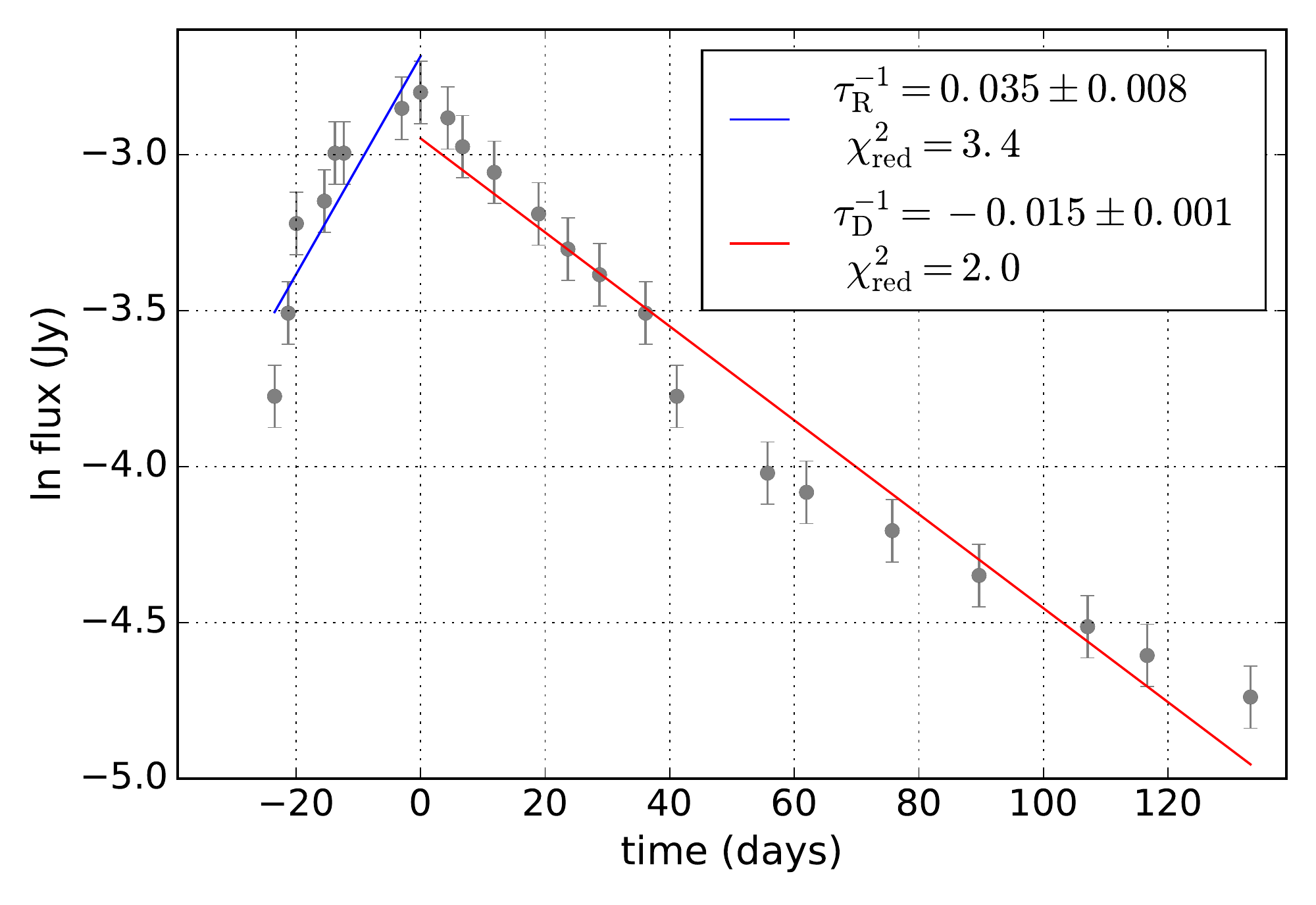}
\caption{Example of a light-curve (RS Oph, \citealt{RSOph}) limited to the flaring event, with no information about the background flux density. }
\label{lits}
\end{figure}

\subsubsection{Long duration flares selected from GBI (14\% of all light-curves; 2\% of analysed flares)}
\label{datagbis}
This group of light-curves consists of long duration ($\approx$~months) flares,
collected from the GBI database. As in the previous group, the datasets did not 
provide enough information for the estimation of the background and the quiescent variability, and required
both of these parameters to be calculated in the same way as described in Section~\ref{datalits}.
However, here the additional difficulty was introduced by scattered noise in the data. 
Some of the noisy data points have been falsely recognised by the AFF as the beginning and/or 
end of the flare. The upper panel of the Figure~\ref{gbis} shows an example of such false detections.
This problem has been solved by smoothing the data, with a window size of 
$\approx$~20 days. This size of the window has been chosen based on the expected  
time-scale of the AGN intrinsic variability. Variabilities in those types of objects,
observed on time-scales shorter 
than 20 days are more likely to be associated with scintillation. 
Lower panel of Figure~\ref{gbis} shows the result of smoothing, with flaring events 
identified by the AFF.

\subsubsection{Pre-selected flares compiled from the literature (41\% of all light-curves; 5\% of analysed flares)}
This group of datasets 
contains mainly single, pre-selected flaring events, with no information about the 
light-curve before and after the outburst. Included here are flares from supernovae, GRBs, classical novae and tidal disruption events. In those cases, we decided 
to measure rise and decline time-scales directly (Figure~\ref{lits}), using method described in Section~\ref{sec:mvt}.\\

For light-curves specified in Sections~\ref{datalits} and \ref{datagbis}, the sigma clipping routine as described in Section~\ref{aff} failed to detect 70~per~cent of the flares due to insufficient information about the background. We note that setting the background level in this alternative way is a subjective choice and can introduce some bias in the analysis. 
Therefore, we have repeated the analysis for a range of parameters, setting the quiescent level to 10, 15, 20 and 30th percentile of the flux density. The results were then investigated 'by eye' to decide which approach selected the highest percentage of flares. While each of those estimations caused some of the flares to be missed, we found that the 15th percentile was the optimal choice for the discussed dataset. Figure~\ref{missed-flare-example} in Appendix~\ref{additional_figures} gives an overview of flares selected from an example light-curve depending on the background estimation. It shows that setting the threshold too low can merge poorly separated flares into one, while too high values results in missing fainter flares. 
We have also checked how different choices of the background in those groups of light-curves affect the final results -- see Figure~\ref{background-est-var} in  Appendix~\ref{additional_figures} and refer to the analysis in Sections~\ref{sec:mvt} and \ref{results}.
Since the discussed datasets form the entire sample of flare stars, as well as most of AGN light-curves, the effects of changing the flare detection threshold are most visible in those two populations of sources (left panel in Figure~\ref{background-est-var}). 
The centre of the diagram, which is dominated by flares that do not suffer the background uncertainty (fast GBI light-curves, individual flares from GRBs, SNe and Novae) shows minimal difference. The right panel of Figure~\ref{background-est-var} demonstrates that the choice of background does not have a significant impact on final probability distributions, due to the affected classes of object populating the extreme ends of the time-scale parameter space.\\
Overall, the automatic selection of flares fails in about 15--20~per~cent of cases, missing flares that can be easily identified by eye. Those are mainly superimposed or close together flares which can not be separated with a simple thresholding method, such as the ones presented in Figure~\ref{background-est-var}.


\subsection{Measuring variability time-scales}
\label{sec:mvt}

We have run the AFF on the whole sample of light-curves adjusted as described in previous section, selecting 1290 single outbursts for further analysis, which, together with 38 pre-selected flares, gives the total number of 1328 flaring events from 86 distinct objects. 

In order to measure rise and decline time-scales for each of the selected flares, we attempted fitting exponential functions to the data: 

\begin{equation}
F = A e^{st}+B, 
\end{equation}

\noindent where amplitude $A$, background $B$ and characteristic time-scale $s$ are the free parameters of the fit.
Exponential fits provide distance independent measurements of the rise/decline rates, and, are physically motivated \citep[][where the flux density of a source is an exponential function of the optical depth.]{vdl}
However, we found that in most cases the exponential fit would fail. One of the likely reasons 
for that is the quality of the data, where time sampling is often insufficient for fitting the three parameters model. 

\noindent Therefore, we decided to estimate the background emission and subtract it from the flare prior to the fit. One obvious choice of this background is the quiescent flux density level $b$ estimated earlier for each of the light-curves (Sections~\ref{aff} and \ref{prep}). However, for poorly sampled light-curves, the beginning/end of the flare (defined as in Section \ref{aff}) can sometimes fall below this background level (for example, the first flare detected in lower panel of Figure~\ref{litm}). Subtracting this background  would result in selecting only part of the flare for further analysis.
In order to make sure that the entire flare is selected, we decided to calculate a separate background level for each flare $b_{\mathrm{flare}}$ either as the first percentile $p^{\mathrm{1st}}$ of the flux values of the light-curve ($F_{\mathrm{all}}$) or the minimum flux density measurement of a given flare ($F_{\mathrm{flare}}$):

\begin{equation}
b_{\mathrm{flare}} = \mathrm{min} \left(p^{\mathrm{1st}}\left(F_{\mathrm{all}}\right); \mathrm{min}\left(F_{\mathrm{flare}}\right)\right).
\end{equation}

\noindent The background emission calculated this way is then subtracted from the flux density measurements of every flare selected by the AFF.

\begin{figure}
\includegraphics[scale=0.42]{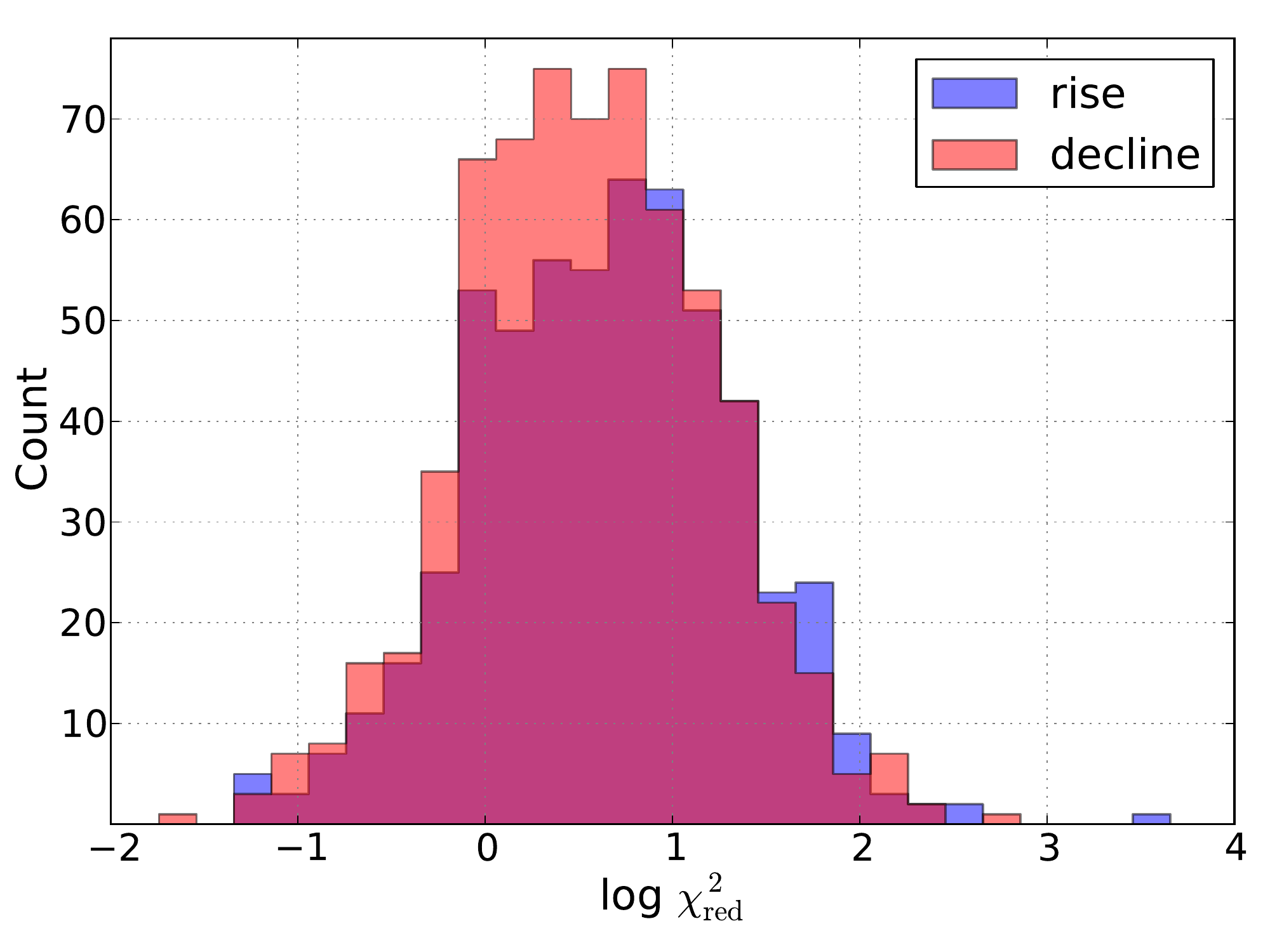}
\caption{Distribution of the $\mathrm{log}\, \chi^{2}_{\mathrm{red}}$ values for rise and decline rates calculated as in Equation~\ref{fiteq}.}
\label{chi2}
\end{figure}

\noindent For each single flare we measure the rise and decline rates, $s$, by fitting the linear 
function in the log-lin space (see Fig~1.~in PFK15): 

\begin{equation}
\mathrm{ln} \left( F - b_{\mathrm{flare}} \right) = st+const.
\label{fiteq}
\end{equation}

\noindent No measurements have been taken for rise/decline phases consisting of three data points or less. Excluding flares for which neither rise or decline phase could be measured decreased the final sample to 804 flaring events, with 564 rise and 649 decline phases fitted successfully. Fits to the data have been done 
using the {\em curve fit} routine from {\em scipy.optimize} package, and include error bars of the flux measurements.
Figure~\ref{chi2} shows the distribution of $\chi^{2}_{\mathrm{red}}$ values for 
measurements of rise and decline phases of the flares. It should be noted that for $\approx$~30~per~cent of measurements the $\chi^{2}_{\mathrm{red}}$ is higher than~10. This might partly be caused by underestimating error bars of flux measurements. However, we expect that the main reason is that the simple two parameter model used in the analysis often fails to accurately fit the data, which may require more complex approach.

\begin{figure*}
\includegraphics[width=\textwidth]{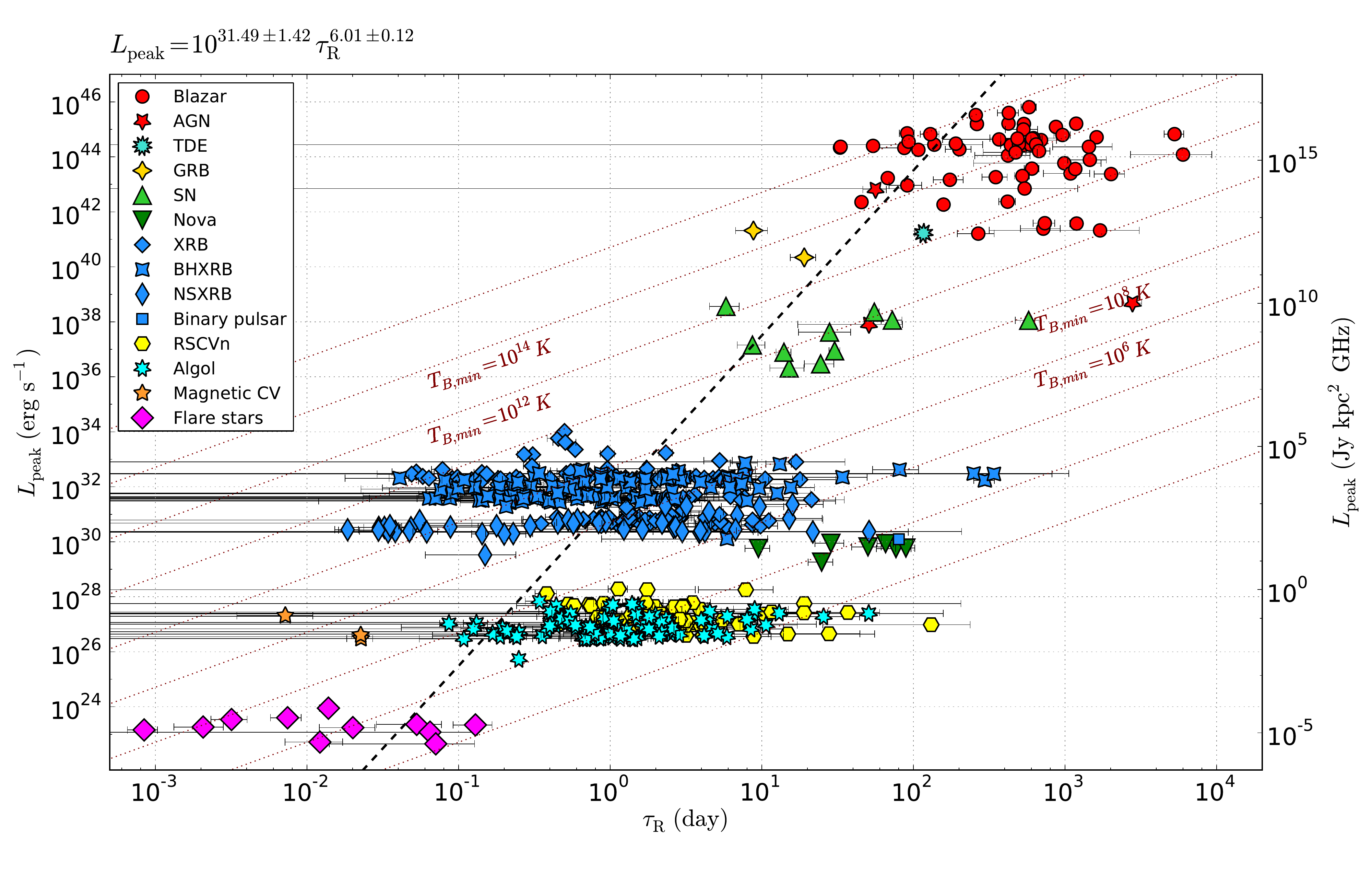}
\caption{Peak radio luminosity plotted against exponential rise time-scale for a range of synchrotron emitting sources, approximately of the form $L \propto \tau^{6}$.}
\label{LT}
\end{figure*}

\noindent Having a set of measurements for each flare, consisting of 
peak flux density, rise/decline time-scales and reduced $\chi^{2}_{\mathrm{red}}$,
we can map those results 
with a corresponding class to which the source belongs, its distance and the 
observed frequency. This enables us to calculate peak radio luminosity of each flare. Results 
of those measurements for the rising phase are shown in Figure~\ref{LT}, where we plot 
peak radio luminosity against rise time $\tau$ of the event (where $\tau=1/s$). The overall correlation of the form 
$L \propto \tau^{6}$ is steeper than $L \propto \tau^{5}$ obtained by manual measurements.
In particular, the relation presented here shows more scatter for relatively lower-luminosity 
classes such as XRBs or RSCVn, than it did in results reported in PFK15. 
This effect is clearly demonstrated in Figure~\ref{TH}, where histograms of the rise rates measured automatically
are overlaid with those corresponding to manual measurements (in case of manual measurements, the level of background emission for each flare has been estimated by eye and subtracted from the flux density measurements). It should be noted, that the scatter observed in time-scales of the different types of object is partly due to a wide range of time-scales covered by individual sources. A variation of Figure \ref{LT} is shown in Figure \ref{single-source-scatter}, where 
single-source measurements for several classes are highlighted.
We have tested whether rejecting measurements with high and/or low $\chi^{2}_{\mathrm{red}}$ values decreases that scatter, however, neither the final range of the time-scales populated by each class or the overall time-luminosity relation have been significantly changed by those constraints. Therefore, we have decided to include all the measurements.
The observed scatter in measured time-scales is discussed in more detail in Section~\ref{discussion} and 
a summary of the manual and automatic results can be found in Table~\ref{manvsauto}.

\noindent In the analysis described in this section, we have automatically reproduced results of the variability 
time-scales measurements reported in PFK15.  
From this point we focus on converting those results into a method which could be used to perform an initial classification of radio transient and variable sources.

\noindent The diagram shown in Figure~\ref{chart} gives an overview of steps included in the analysis described in this section, as well as a list of remaining steps which are discussed in Sections~\ref{se} and \ref{results}.

\begin{table}
\caption{Parameters of the fits to the rise and decline phases (with the following formula: $\log(L_{\mathrm{peak}}) = a \times \log\tau + b$), for both manual (a$_\mathrm{M}$, $b_\mathrm{M}$) and automatic (a$_\mathrm{A}$, $b_\mathrm{A}$) methods. }
\begin{threeparttable}[b]
\begin{tabular}{|l|@{}c|@{}c|c|@{}c}
\hline
& a$_\mathrm{M}$ $\pm \delta$ a$_\mathrm{M}$ & b$_\mathrm{M}$ $\pm \delta$ b$_\mathrm{M}$ & a$_\mathrm{A}$ $\pm \delta$ a$_\mathrm{A}$ & b$_\mathrm{A}$ $\pm \delta$ b$_\mathrm{A}$  \\
\hline 
Rise        & 5.21 $\pm$ 0.15 & 31.08 $\pm$ 1.84 &  6.01 $\pm$ 0.12  & 31.49 $\pm$ 1.42 \\
Decline & 5.09 $\pm$ 0.22 & 29.44 $\pm$ 2.88 &  6.70 $\pm$ 0.22  &  29.11 $\pm$ 2.11 \\
\hline
\end{tabular}
\end{threeparttable}
\label{manvsauto}
\end{table}

\section{Sky Densities}
\label{se}

We aim to use the results obtained in Section~\ref{sec:mvt} as a base for the probabilistic classification method by modelling the measured time-scale distributions as Gaussian functions, while assuming that for each class we have reasonably well measured the width of the distribution.
However, the compiled dataset consists of observations originating from various surveys, 
performed with a wide range of instruments with different sensitivities, fields of view and other parameters which contribute to certain preferences in the available sample of light-curves. As such, the analysed dataset does not accurately represent the underlying populations of considered classes of object.
Although it is not possible to properly account for all these biases, in this section we attempt to estimate the areal densities of objects within each class. This correction, although basic, provides more realistic picture of time-scale distribution for different populations of radio transient and variable sources.
\\

\subsection{Method of estimating areal densities}
\label{sec:method}

{\em Extragalactic events.} Areal density estimates for most of the extragalactic classes of objects 
are available in the literature. 
Here we extrapolate these results to our chosen flux density limit 
of 0.1~mJy, assuming that events of the extragalactic origin are distributed as:

\begin{equation}
N \propto S^{- \frac{3}{2}},
\end{equation}

\noindent where $N$ is the expected number of sources observed with the 
flux density limit $S$.

\noindent{\em Galactic objects.} Space densities of Galactic objects have been
studied previously 
and are available in the literature.
In order to convert spatial distribution into areal density, we need
to consider the volume of the Galaxy each class populates. First,
we calculate the maximum distance at which source belonging to
considered class can be observed, with assumed flux density limit.
We estimate a typical radio luminosity of the class and use the 
following relation to evaluate the distance:

\begin{figure}
\includegraphics[scale=0.54]{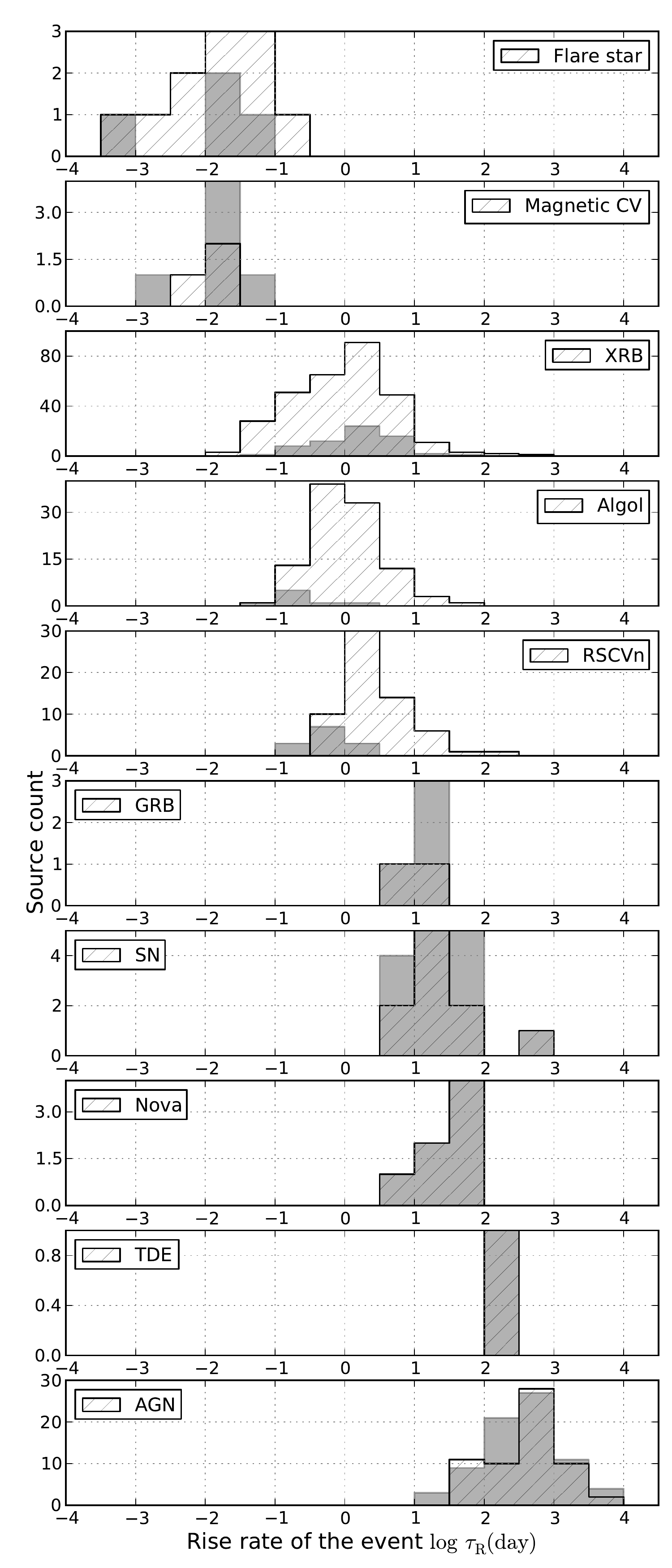}
\caption{Dashed histograms represent distribution of automatically measured rise rates for a range of classes of objects. For comparison, overplotted in grey are previously reported \citep{paper1} manual measurements.}
\label{TH}
\end{figure}

\begin{figure}
\includegraphics[scale=0.31]{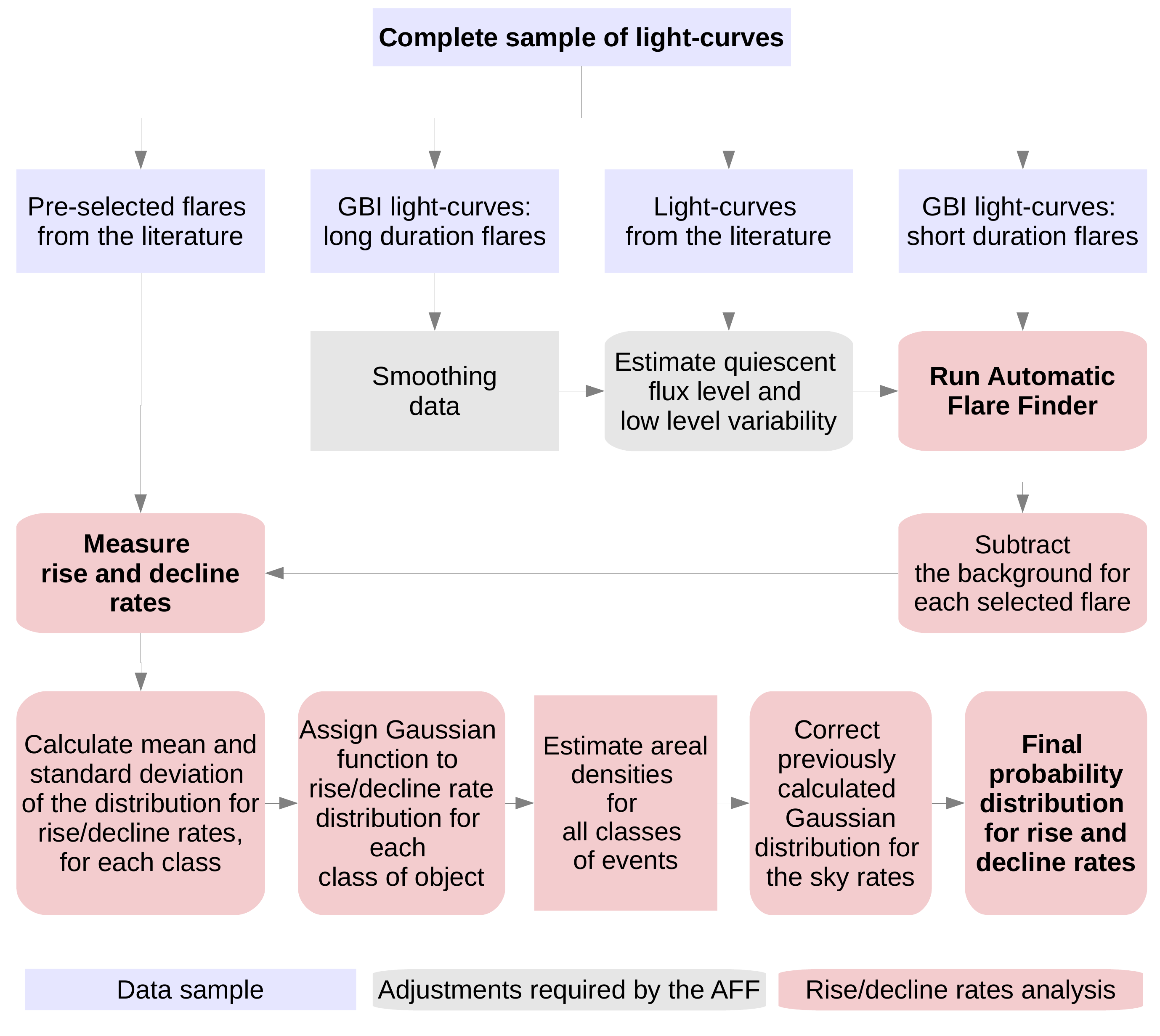}
\caption{Illustration of the process flow showing the main steps of the analysis presented in Section~\ref{analysis}, from the compiled sample of the data to measuring rise/decline rates. The remaining steps, listed in lower panel, are described in Sections~\ref{se}~and~\ref{results}. Rectangular frames correspond to parts of the analysis done manually; automatic steps are listed in rounded frames.}
\label{chart}
\end{figure}

\begin{equation}
d_{\mathrm{max}} = \sqrt{\frac{L}{4 \pi F_{\mathrm{lim}}}} .
\end{equation}

Unlike for extragalactic sources, the space density $\rho$ of galactic sources depends on their position in the Galaxy and drops as we move out of the Galactic plane and away from the Galactic center. 
The density profile can be described as

\begin{equation}
\rho(z) = \rho_0 \mathrm{e}^{\left(-|z|/h\right)},
\end{equation}

\noindent where $z = d \mathrm{sin}b$ is the distance from the Galactic plane ($d$ is distance to the source, $b$ is the Galactic latitude), $h$ is the scale height (describing a distance over which the number of sources drops by factor of $e$) and $\rho_0 = \rho(0)$ is the local space density of objects (we assume that the radial dependence of $\rho$ is negligible). With this assumption  we calculate the volume of space occupied by a given class of objects within distance $d$ (details of the calculation of generalised volume of space depending on the density profile can be found in  \citealt{vmax}). 
Finally, including a duty cycle estimate for each class of object gives the 
actual number of sources visible in the sky in a single snapshot. 
Table~\ref{rate} summarises areal densities evaluated for each class.
Details of the calculations are described in the following sections.

\subsection{Types of transients and variables}
\label{tt}

\subsubsection{Tidal Disruption Events}
\label{tde}

There have been several radio detections of tidal disruption events (TDEs) to date, however, only two of them are sampled well enough to be included in this analysis.
\citet{Frail2012} estimated the sky rate of such events to be 
0.1~deg$^{-2}$ at the 0.3~mJy flux density limit. Using the method described in 
Section~\ref{sec:method} we estimate that at the 0.1~mJy limit, 
the TDE rate is $\approx$~0.52~deg$^{-2}$.

\subsubsection{Gamma Ray Bursts}
\label{grb}

The areal density of Orphan GRB Afterglows -- classical GRBs with the explosion 
axis directed away from the line of sight, which should dominate, is derived by \citet{Frail2012} 
at the 0.3~mJy flux limit for a 10$^{\circ}$ beam opening angle. 
Their result of 0.01~deg$^{-2}$ scaled down to 0.1~mJy gives 0.052~deg$^{-2}$ in a single snapshot.

\subsubsection{Supernovae}
\label{sn}

\citet{Frail2012} give areal densities for SNII, SNIbc and SN~1998bg-like sources. Adding all those sky 
densities and extrapolating the results to our flux density limit, gives a rate of 0.21~deg$^{-2}$ at a single epoch.

\subsubsection{Active Galactic Nuclei}
\label{agn}

In order to estimate the surface density of radio-variable AGN, we looked at the number of all variable objects found in the FIRST
survey data. Out of 1600 variable sources reported by \citet{FIRST}, 489 were identified to be AGN and
variable galaxies. Another 120 are believed to be highly variable quasars. The analysed observations 
covered 8444~deg$^{2}$, at the sensitivity of 0.15~mJy. The source detection threshold 
chosen for the analysis was 1~mJy. Extrapolating this result to 0.1~mJy with the spherical approximation 
described in Section~\ref{sec:method}, gives the AGN sky density of 2.3~deg$^{-2}$ at single epoch.

\subsubsection{X-ray binaries}
\label{xrb}
\noindent There are about 1000 X-ray binaries within our Galaxy, which, when above the flux density limit of 0.1~mJy chosen in this work, are in outburst and can be detected in a radio survey \citep[e.g.][]{xrbrate}. With the estimated duty cycle of XRBs of $\approx$~1\% we get $\approx$~10 sources in the sky, at a single epoch. This gives a rate of about 2.4~$\times$10$^{-4}$~deg$^{-2}$ objects at 0.1~mJy flux density limit.

\subsubsection{Non-magnetic CVs}
\label{dn}

\noindent {\em Dwarf Novae}. Having only one radio light curve of a Dwarf Nova (SS Cyg), 
we use the peak radio luminosity measured for that source in order to estimate 
the maximum distance to which objects of this class can be detected at 0.1~mJy flux density limit. Following the steps described in Section~\ref{sec:method} we get d$_{\mathrm{max}}$~=~380~pc. \citet{PK2012} derive space density of non-magnetic 
cataclysmic variables to be 4~$\times$~10$^{-6}$~pc$^{-3}$ with the scale height of 260~pc. The estimated 
duty cycle of Dwarf Novae is 1-15~per~cent (Servillat et al. 2011) -- here we assume it is $\approx$~10~per~cent. 
Applying the method discussed in Section~\ref{sec:method} we get the areal 
density of 0.0013~deg$^{-2}$.\\

\noindent{\em Nova-likes}. Nova-like CVs were until recently not thought to be radio sources. However, \citet{kording2011} and \citet{coppejans2015} have now shown that several of these systems are radio sources at a level of less than about 0.2~mJy, with variability detected in two systems. Nova-likes will therefore show up as variable sources in sensitive radio surveys. We will not consider them here, however, because we do not expect to see many of them as transients. The reasons are their low space density (more than an order of magnitude less than DNe) and the fact that Nova-likes do not show frequent large amplitude variability. 

\begin{table*}
\caption{Estimated areal densities of the studied classes of object at 0.1~mJy flux density 
limit, together with the mean and standard deviation parameters calculated for each class and used to estimate their Gaussian distribution, for both rise and decline phases.}
\centering
\begin{tabular}{p{2cm}p{1.5cm}p{7cm}p{1cm}p{1cm}p{1cm}p{1cm}}
\hline
Class & Rate (deg$^{-2}$) & References & log$\,\mu_{\mathrm{R}}$ (day) & log$\,\sigma_{\mathrm{R}}$ (day)& log$\,\mu_{\mathrm{D}}$ (day) & log$\,\sigma_{\mathrm{D}}$ (day) \\
\hline
AGN & $2.3$  & \citet{FIRST} & 2.6 & 0.51 & 2.57 & 0.48 \\
TDE & $0.52$  & \citet{Frail2012}& 2.07 & 0.46 & 2.84 & 0.25\\
GRB afterglow & $0.052$ & \citet{Frail2012} & 1.11 & 0.17 & 1.94 & 0.46\\
SN & $0.21$ & \citet{Frail2012} & 1.47 & 0.54 & 2.43 & 0.68 \\
SGR & $10^{-8}$ & \citet{Olausen2014}, \citet{Ofek2007} & -- & -- & 0.68 & 0.43\\
XRB & $2.4 \times 10^{-4}$  & \citet{xrbrate} &  0.0044 & 0.73 & 0.07 & 0.69 \\
Dwarf Nova & $0.0013$ & \citet{PK2012}, \citet{Servillat2011} & --  & -- & 0.92 & 0.43\\
Classical nova & $0.0023$ & \citet{Roy2012} & 1.60 & 0.32 & 2.39 & 0.35\\
RSCVn & $0.011$ & \citet{Williams2013},  \citet{Favata1995}, \citet{Ottmann1992} & 0.45  & 0.47 & 0.46 & 0.45\\
Algol & $0.062$ & \citet{Duerbeck1984} & 0.031 & 0.51 & 0.15 & 0.50\\
Flare star & $0.0064$ & \citet{Reid2007}, \citet{Osten2008}  & -1.85 & 0.67 & -1.86 & 0.43\\
Magnetic CVs:  &  &   & &  & & \\
Polars & 0.038 &  \citet{polarsdc}, \citet{Pretorius2013} & -- & -- & -- & -- \\
IPs  &  0.0008 & \citet{ipspacedens}, \citet{Pretorius2013} & -- & -- & -- & --   \\
Polars + IPs & 0.039 &  -- &  -1.81 & 0.23 & -1.93 & 0.069 \\
\hline
\end{tabular}
\label{rate}
\end{table*}

\subsubsection{RSCVn}
\label{rscvn}

The space density of RSCVns  is approximately 6 $\times$ 10$^{-5}$ pc$^{-3}$ \citep{Favata1995}. The duty cycle of the class is estimated to be $\approx$ 10 per cent and the typical monochromatic radio luminosity of sources 10$^{16}$ erg s$^{-1}$ Hz$^{-1}$ \citep{Williams2013}. Converting space density into areal density as in Section~\ref{sec:method}, with the scale height of 325 pc for RSCVns \citep{Ottmann1992}, we get the sky density of 0.011 deg$^{-2}$.\\

\subsubsection{Algol binaries}
\label{algol}

We use a value of $\approx$~5$\times$10$^{-6}$~pc$^{-3}$ for the space density of algol binaries, as defined by \citet{Duerbeck1984}.
To calculate the maximum distance of detection at the 0.1~mJy flux density limit we use the average monochromatic luminosity 
of objects in our sample (1.3$\times$10$^{17}$~erg~s$^{-1}$~Hz$^{-1}$). There are no good estimations of duty cycle for algol binaries -- 
based on the light-curves of sources in our sample we estimate it to be $\approx$~23~per~cent. 
Correcting for the scale height of algols, which is $\approx$~400~pc \citep{Duerbeck1984}, we get the 
final areal density of 0.062~deg$^{-2}$.

\subsubsection{Novae}
\label{cn}

The rate of Novae in the Milky Way is estimated to be $\approx$~35~year$^{-1}$ \citep{Roy2012}. 
In order to calculate the maximum distance at which a nova could be detected, with a limiting flux density of 0.1~mJy, we assume that the typical radio luminosity it can reach in an outburst is $\approx$~1.2~$\times$~10$^{20}$erg~s$^{-1}$~Hz$^{-1}$, that is, the average peak radio luminosity of novae in the sample. 
Because the derived distance is 30~kpc, we can estimate that approximately all 35 novae per year in our Galaxy could be 
detected at the assumed flux limit. The final rate in a single snapshot is evaluated by multiplying a 
rate per year by the duration of nova. Based on the properties of light curves in our sample, 
we expect to detect a typical nova in the outburst for $\approx$~1000 days at 5-8~GHz frequency. 
For the parameters listed above, we estimate that the areal density of novae is 0.0023~deg$^{-2}$.

\subsubsection{Soft Gamma-ray repeaters}
\label{sgr}

\noindent The sky density of SGRs is not well constrained. \citet{Ofek2007} gives the upper limit for the rate of giant flares (similar to SGR~1806-20 event) of 5~$\times$~10$^{-4}$~year$^{-1}$ per SGR.
With the number of SGR sources in the Galaxy reported by \citet{Olausen2014}  of 15\footnote{\url{http://www.physics.mcgill.ca/~pulsar/magnetar/main.html}}, and the duration of the radio flare analysed in our sample ($\sim$~20~days) we get the areal density of approximately 10$^{-8}$~deg$^{-2}$.\\

\subsubsection{Magnetic CVs}

\noindent {\em Polars}. One of the two magnetic CVs in our sample -- V834 Cen -- belongs to the subclass of polar magnetic CVs.
With the luminosity measured in the analysis (2.4 $\times$ 10$^{17}$ erg s$^{-1}$ Hz$^{-1}$), we estimate the maximum distance of $d_{\mathrm{max}}$~=~1.4~kpc.
For the space density of 10$^{\mathrm{-6}}$~pc$^{-3}$ \citep{Pretorius2013}, duty cycle of 0.5~\citep{polarsdc} and the scale height equal to 260~pc \citep{Pretorius2013}, we estimate the sky density for these types of objects of 0.038~deg$^{-2}$.\\

\noindent {\em Intermediate polars}. Sky density of IPs is calculated based on the AE Aqr.
With the average measured luminosity of that source (5.8~$\times$~10$^{16}$~erg~s$^{-1}$~Hz$^{-1}$), the estimated maximum distance for the detection is 
d$_{\mathrm{max}}$~=~700~pc. 
With space density of 10$^{-7}$~pc$^{-3}$ \citep{ipspacedens}, scale height 120~pc \citep{Pretorius2013} and duty cycle of 1 (persistent sources) we get the estimated sky density of 0.0008~deg$^{-2}$.

\subsubsection{Flare Stars}
\label{fs}
Space density of flare stars given by \citep{Reid2007} is 0.08~pc$^{-3}$. The average luminosity of flare stars in our sample is 4.7$\times$10$^{13}$~erg~s$^{-1}$~Hz$^{-1}$. 
With such low luminosities, at the 0.1~mJy flux limited survey they can be detected up to a distance of $d_{\mathrm{max}}~\approx$~20~pc. Within that distance, which is relatively small compared
to the size of the Galaxy, we can assume a uniform, spherical distribution of sources. The duty cycle of flare stars is not well constrained (\citealt{Osten2008}, \citealt{Hilton2010}) -- assuming the upper limit of 10~per~cent, we get the areal density of 0.0064~deg$^{-2}$.

\noindent Flare stars can also produce bright coherent bursts, which at the sub-milijansky detection threshold can be detected from a distance of several hundred parsecs \citep{Osten2008}. 
Although these types of events are quite rare (duty cycle less than 1 per cent), because of their high brightness, the areal density of coherent flares can reach approximately 1.2~deg$^{-2}$ at the 0.1~mJy flux density limit. On the other hand, they are expected to evolve rapidly ($\sim$~60~s) and should only be common in surveys exploring very short time-scales.\\

\section{Probability distribution of variability time-scales}
\label{results}

For a transient/variable candidate discovered in a blind radio survey, it is very unlikely that the 
information about its distance, and therefore the luminosity, will be known (at least immediately). 
Thus, all of the earliest information will be contained in the variability of the 
light-curve and spectral distribution of the source. Correspondingly, if we can not 
take into account information about the luminosity of the source, the relation shown in 
Figure~\ref{LT} is reduced to a one dimensional histogram of variability time-scales (Figure~\ref{TH}), 
increasing the uncertainty in separation between different classes of objects. This result, however, can be 
converted into the probability distribution and convolved with the expected areal densities 
of radio sources. 

In order to represent the variability time-scales as a probability distribution, 
for each of the considered classes of object we have calculated the mean value~$\mu$ 
and the standard deviation~$\sigma$ (listed in Table~\ref{rate}) of the variability time-scales~$\tau$. 
Using these values we have assigned the following 
Gaussian probability distribution to each class:

\begin{equation}
P(\tau) = \frac{1}{\sqrt{2 \pi } \sigma} e^{-\frac{(\tau - \mu)^{2}}{2 \sigma^{2}}}.
\label{rawprob}
\end{equation}

\begin{figure*}
\includegraphics[scale=0.5]{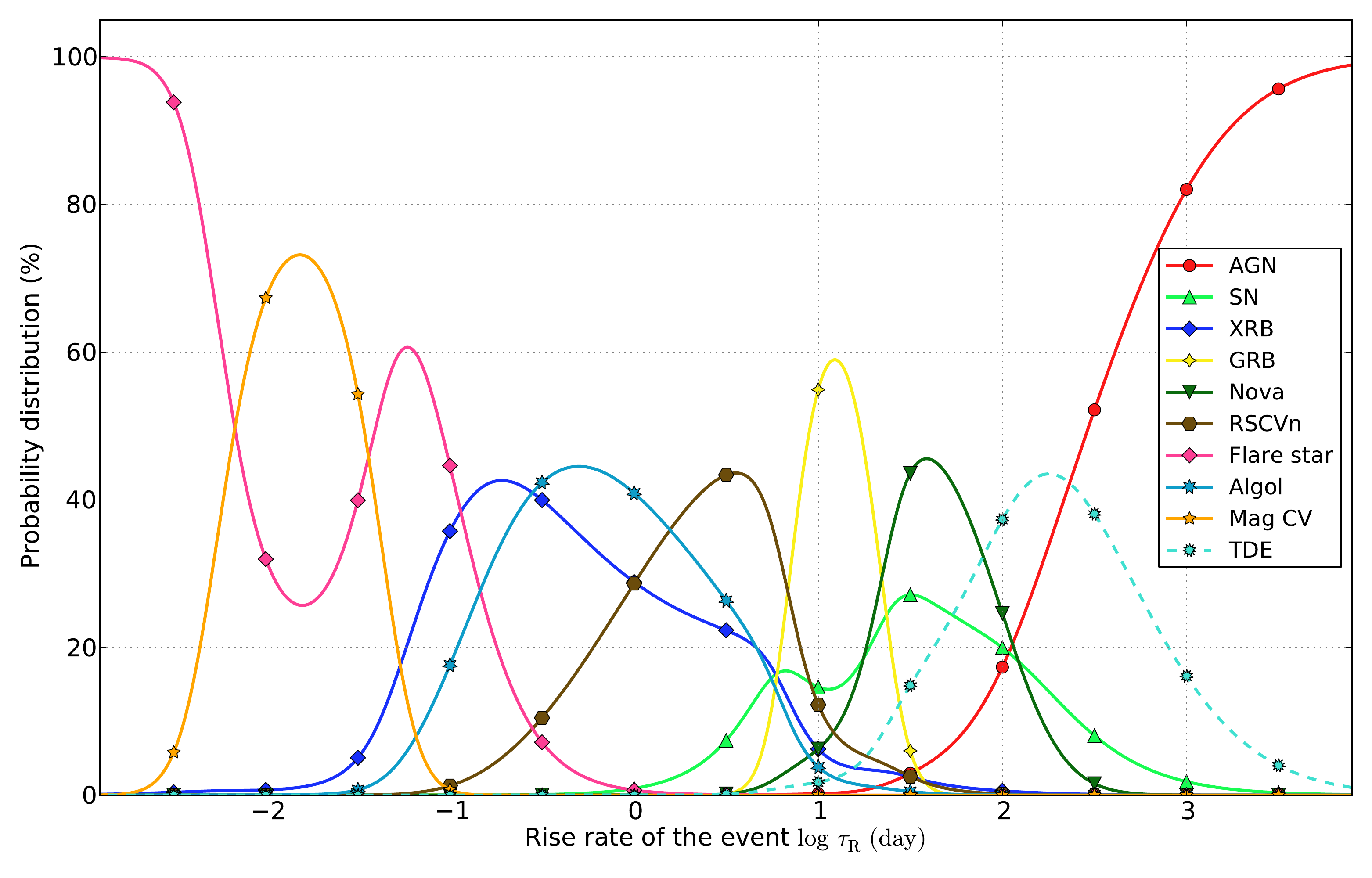}
\vspace{0.6cm}\\
\includegraphics[scale=0.5]{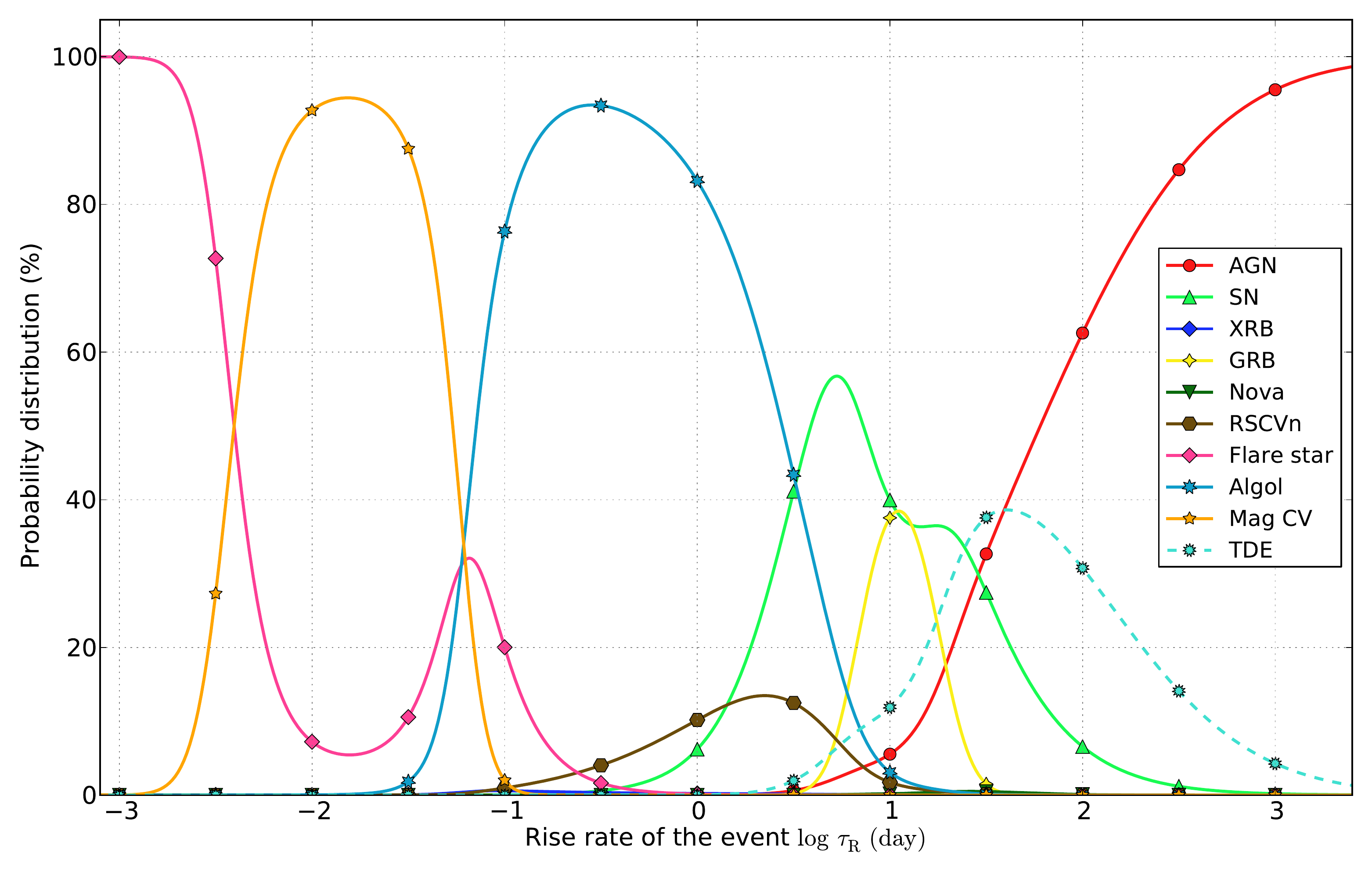}
\caption{ {\em Upper:} Relation shown in Figure~\ref{LT} converted into the probability distribution of source belonging to given class of objects based on the rise time-scale, normalised to 100 per cent. {\em Lower:} The same relation, convolved with the estimated sky densities of objects (Table~\ref{rate}). These estimated distributions have been calculated with the flux density limit of 0.1~mJy.}
\label{RP}
\end{figure*}

\noindent The upper panel in Figure~\ref{RP} shows the probability distribution of all the classes of objects in our sample, 
normalized to 100~per~cent. The dashed line represents the distribution of TDEs, for which only one flare with well sampled rise time was available. In this case, the mean value of the 
distribution has been approximated by the variability time-scale measured for that source, 
while the standard deviation has been calculated as the mean value of the
standard deviations measured for all the other classes of objects. 
The probability distribution obtained in Equation~\ref{rawprob} depends strongly on biases in our sample. 
In order to convert it into the actual probability distribution, which takes into account expected number of sources in the sky, 
we multiply the distribution obtained in Equation~\ref{rawprob} by the areal density of objects $R$, estimated in Section~\ref{se}

\begin{equation}
P_{\mathrm{corr}}(\tau) = R \times P(\tau).
\end{equation}

Given a small number of light-curves representing each subclass of Magnetic CVs, we add
their sky densities and calculate their combined probability distribution instead of considering them separately.
With this approach, we avoid adding the uncertainties associated with assigning Gaussian distribution based on approximated parameters.
The lower panel in Figure~\ref{RP} shows the probability distribution of all the types of objects in our sample, corrected
for the estimated sky rates of different classes. It is worth noting, that sources like XRBs or Novae, even though they make up a significant fraction of the original sample, due to 
their low expected sky densities, are very unlikely to be found in a blind radio 
transient survey. 

The lower panel in Figure \ref{RP} provides a template of the variability time-scales which could be used as an indicator of the newly discovered sources class.
At the moment, the number of available flares used in designing this classification 
method is uneven across a range of objects included. However, it can be gradually 
improved with new sources detected in future surveys, where additional data could help to limit the uncertainties of the initial classification.
This would be especially desirable in the case of classes which have few well-studied light-curves to date, such as tidal disruption events, for which at this point, the variability time-scales are not well constrained.
Tables~\ref{probR} and  \ref{probD} give the probability information for each of the classes of objects 
(for rise and decline phases respectively) on a time-scale $\approx$~10$^{-4}$--10$^{4}$~days, with a logarithmic time-step of~0.5.

The presented results depend on a range of parameters, which we can not accurately account for in our analysis.  Although we have corrected for the expected areal densities of objects down to 0.1~mJy, for higher sensitivity surveys intrinsically faint sources will become progressively more common, with the number of Galactic sources increasing up to the point where the entire population within the Galaxy can be detected.
The form of the probability distributions which can be applied to a given survey will also depend on the observing frequency, which might favour detection of sources showing certain variability time-scales, and -- for Galactic populations -- pointing location. At the moment, these have to be considered separately within a given survey, but, in the future might be included as part of a more advanced pipeline.

\section{Discussion}
\label{discussion}

Figure~\ref{TH} shows that creating a simple automatic flare finder allowed us to include a larger number of flares 
in the analysis, compared with the manual selection.
Although there is an overall agreement between both methods, 
comparison of results obtained 
automatically (Figure~\ref{LT}) and manually (Figure~3, PFK15) shows a higher 
degree of scatter for several classes of objects in the automatic approach.
Light-curves corresponding to these sources (XRB, RSCVn, Algol) consist of tens of single outbursts, 
which in the manual analysis we have limited to those with the best data quality.
The flare finding software, however, had identified all of the flaring 
events in the same light-curves, increasing the range of obtained time-scales for 
each source. Although most of the measurements correspond to real flares,
in marginal cases a false detection has also been included. 
For example, an odd noisy data point within a flare can be mistaken for the end  
of the respective flare and result in underestimating measured decline time. In other extreme cases, 
several outbursts superimposed such that the flux level does not drop low enough 
for the AFF to separate them, will be measured as one long flare rather than single events, 
and make the rise/decline times appear longer than they really are. 
Because several of those exceptions identified in our analysis could not be
removed other than by manually excluding them, we have decided to leave them in the final 
results in order to keep the process purely automatic. In this way we also begin to learn what pitfalls might befall a future automated system.

\begin{figure}
\includegraphics[scale=0.2465]{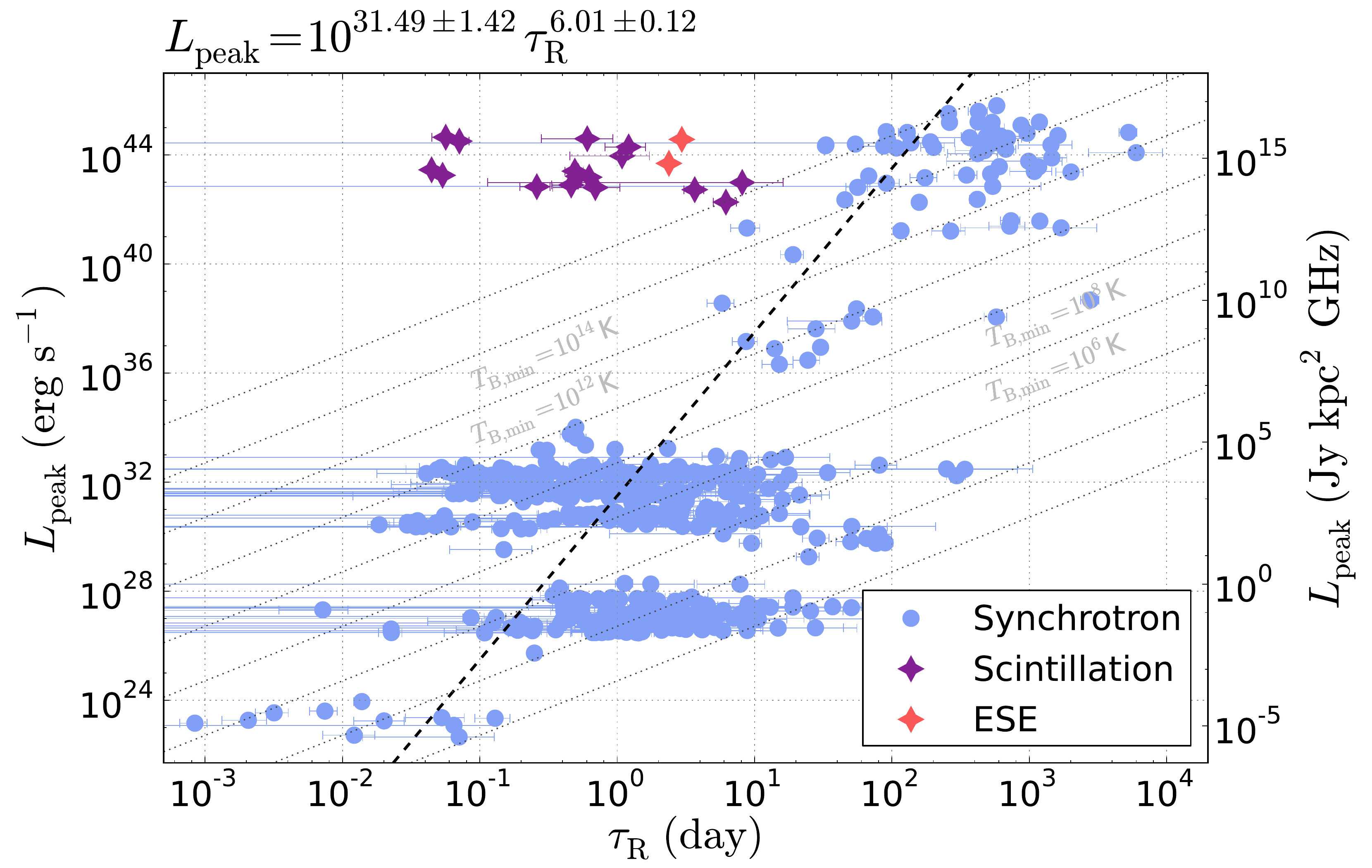}
\caption{Variation of Figure \ref{LT}, where all events originating from synchrotron emission are marked with blue circles, and additional measurements corresponding to scintillating sources and extreme scattering events (ESEs) are plotted in purple and red respectively. As discussed in Section \ref{discussion}, scintillating sources together with extreme scattering events appear to populate most of the time-scale parameter space and cover the range of rise rates typical for all the classes, from low luminosity stars to AGN. Scintillation and ESEs have not been included in the fit.}
\label{mainsct}
\end{figure}

This work is exploring the parameter space typical for flares originating from incoherent emission processes.
Variability from sources of coherent emission such as pulsars is usually detected in high-time resolution observations, and, with duration of milliseconds--seconds, is outside of the time-scales regime considered in this analysis.
However, \citet{coppejans2015} showed that a flare from nova-like source TT Ari, evolving on time-scale of minutes, is most likely produced by a coherent emission mechanism. The characteristic exponential rise time of that flare ($\approx$~0.005~day) shows that in some cases, coherent flares can overlap the time-scales of incoherent events.

One of the main challenges we might come across while
classifying a newly discovered source by its variability rates can be scintillation.
Figure~\ref{mainsct} shows the relation between the time-scales and peak radio luminosities for our extended sample of objects, where a sample of scintillating sources and extreme scattering events are plotted along with the previously shown (Figure~\ref{LT}) synchrotron events. It shows that the parameter space of rise rates for these types of 
sources overlaps strongly with most of the analysed astrophysical flares. It might be 
possible to distinguish between those based on their spectral properties -- 
while the amplitude of synchrotron flares is lower at lower frequencies, \citet{sctexp} show that the observed variability due to interstellar scintillation
can be stronger at lower frequencies.
Similarly, if originating from intrinsic variability, the light-curve should evolve on longer time-scales at lower frequencies, which is not the case for ESEs, where the peak flux density is observed simultaneously at all wavelengths \citep{ESE1}.
Another way of separating discussed types of events lies in the shape of the flare -- those originating 
from astrophysical explosions are usually characterised by longer decaying times 
comparing to their rise times, while in the scintillation both phases are more symmetrical. 

Although the entire process aims at automating the rise/decline rate measurements,  several steps have been done `by eye' (listed in rectangular frames in Figure~\ref{chart}). Those include splitting the complete sample of light-curves into four distinct groups (each of which required different method of estimating the quiescent flux level) and applying smoothing for one of them. The remaining steps (listed in rounded frames in Figure~\ref{chart}), such as selecting flaring events from the data and measuring rise/decline rates (including background subtraction) have been done automatically.
It should be noted that all the manual adjustments were only required due to the nature of the compiled sample. 
The flare finding routine applied to the dataset was necessarily simple,  however, even this unsophisticated thresholding approach needed to be somewhat flexible in order to work on a variety of available light-curves.
In the future transients/variables surveys the observations carried out with an individual telescope will provide more consistent datasets, and the flare identification method can be replaced by more advanced techniques, such as those discussed in Section~\ref{flarefinding}. 
 Additionally, as the aim of presented classification method is to give an earliest indication of the sources nature, a potential flare should be identified as soon as the variability above the quiescent flux level arises. With the long term monitoring of the background prior to outburst, it is possible that an iteration of a presented flare finding routine could be used as part of a transients/variables pipeline, however, the exact form of the software will depend on the surveys design.

At the moment the analysis is based on a fairly limited sample of radio flares and an incomplete representation of certain types of sources. 
Including more data could restrict the individual distributions and reduce the resulting overlap, providing clearer classification. 
Alternatively, it can not be ruled out that extending the number of flares will lead to even greater scatter of measurements within respective classes, decreasing the separation of their probability distributions. 
At this point, we can not definitively say which of these scenarios is more likely. Gradually updating presented result with new sources might help to provide a more conclusive distribution.

\section{Conclusions}
\label{conclusions}

This work builds upon the previously reported result, which showed a clear correlation 
between the luminosity and time-scale of variable radio objects, ranging from nearby, 
intrinsically faint to distant and luminous sources (PFK15).
Here we proposed a method of classifying radio transient and variable sources using the rise and/or decline rates of their flares.
It is based on the analysis of existing light-curves of synchrotron events, where the individual flares were selected automatically in order to reduce bias associated with choosing data by hand and make the result easy to reproduce.
We convolved the measurements of rise and decline time-scales with the expected areal densities of sources down to a 0.1~mJy flux density limit, providing probability distributions of variability time-scales for a wide range of classes.
We have shown that comparison of the variability time-scales measured for a transient/variable source with the presented distribution could point -- with a given probability -- to its nature. 
The result shows that the variability-based classification can be used -- to some extent -- to differentiate between classes of objects. 
We have also investigated time-scales of  variability caused by propagation effects such as scintillation and extreme scattering events. We showed that these time-scales partly overlap with the parameter space of synchrotron events and might require further analysis to separate from astrophysical flares.

In the future, this method or a derivative could be incorporated into automated transient detection
pipeline \citep[e.g. ][]{trap} and used as part of a more complex detection/classification software by setting initial constraints 
on the class of the newly discovered source. 
The number of transient and variable sources found in future SKA surveys could reach up to 1000 per week \citep{Burlon2015, Fender2015, Perez-Torres2015}, hence,  the initial, time-scale based classification might be valuable in the process of selecting interesting events.
This early indication could then be further verified and narrowed down 
by including additional information such as an optical flux measurement (Stewart et al. {\em submitted}), location in the sky or comparison with the archival data.

\section*{Acknowledgements}
We thank the referee for helpful comments and suggestions which helped to improve this paper.
This project was funded by European Research Council Advanced Grant 267697 
"4 PI SKY: Extreme Astrophysics with Revolutionary Radio Telescopes". 
The Green Bank Interferometer is a facility of the National  Science Foundation 
operated by the NRAO in support of NASA High Energy Astrophysics programs. 
This research has made use of data from the MOJAVE database that is maintained 
by the MOJAVE team (Lister et al., 2009, AJ, 137, 3718). This research has made use 
of the NASA/IPAC Extragalactic Database (NED) which is operated by the Jet Propulsion 
Laboratory, California Institute of Technology, under contract with the National Aeronautics 
and Space Administration.

\appendix

\section{Source tables and data}
Tables \ref{probR} and \ref{probD} will be maintained for download at 4pisky.org and updated as new data come in.


\begin{table*}
\caption{Additional tidal disruption event included in the analysis}
\centering
\begin{threeparttable}[b]
\begin{tabular}{cccc}
\hline
Source & Distance [Mpc] &  Reference & Distance reference \\
\hline
ASASSN-14li &  90 &  \citet{14li} & \citet{14li}\\
\hline
\end{tabular}
\end{threeparttable}
\label{tdesample}
\end{table*}

\begin{table*}
\caption{Sample of scintillating sources included in the analysis}
\centering
\begin{threeparttable}[b]
\begin{tabular}{cccc}
\hline
Source & Distance [Mpc] &  Reference & Distance reference \\
\hline
0917+624 & 10581 & \citet{0917+624} & MOJAVE\tnote{1} /NED\tnote{1} \\
0405-385	& 9151.6 &  \citet{0405-385} & NED \\
1257-326	& 8895.6  & \citet{1257-326} & NED \\
1328+6221 & 8573 & \citet{1328+6221}  & NED \\
1819+3845 & 3141.8 & \citet{1819+3845}  & \citet{1819+3845D}\\
J0510+1800 & 2297.8 & \citet{KoaySCT} & NED \\
J0958+6533 & 1951 & \citet{KoaySCT} & MOJAVE/NED \\
J1734+3857 & 6452.6 & \citet{KoaySCT} & NED \\
J1128+5925 & 13846.4 & \citet{J1128+5925} & NED \\
1144-379 & 7103.1 & \citet{1144-379} & NED \\ 
J0102+5824 & 3843 & \citet{Lovell2003} & MOJAVE/NED\\
\hline
\end{tabular}
\begin{tablenotes}\footnotesize
\item [1] http://www.physics.purdue.edu/astro/MOJAVE/index.html
\item [2] http://ned.ipac.caltech.edu/ 
\end{tablenotes}
\end{threeparttable}
\label{sctsample}
\end{table*}

\begin{table*}
\caption{Sample of extreme scattering events included in the analysis}
\centering
\begin{threeparttable}[b]
\begin{tabular}{cccc}
\hline
Source & Distance [Mpc] &  Reference & Distance reference \\
\hline
0954 + 658 & 1951 & \citet{ESE1} & MOJAVE/NED\\
AO 0235+164	&  6142 &  \citet{ESE2} & MOJAVE/NED \\
\hline
\end{tabular}
\end{threeparttable}
\label{esesample}
\end{table*}

\begin{table*}
\caption{Probabilities distribution of variability time-scales for rising phase of the flare, 
for a range of measured rise rates (time given on logarithmic scale). These distributions have been calculated for the flux density limit of 0.1~mJy.}
\centering
\begin{threeparttable}[b]
\begin{tabular}{ccccccccccc}
\hline
 log $\tau$ (day) & AGN& SN&XRB&GRB&Nova&RSCVn&Flare stars&Algol&Magnetic CV&TDE	 \\
\hline
-4.0	  	\hspace{0.1cm}	  	-3.5	 & 	4.3e-26	 & 	2.6e-15	 & 	0.00041	 & 	4.9e-162	 & 	2.3e-54	 & 	4e-13	 & 	100.0	 & 	3.5e-07	 & 	1.7e-08	 & 	6.8e-28\\ 
 -3.5	  	\hspace{0.1cm}	  	-3.0	 & 	5.6e-22	 & 	2.1e-12	 & 	0.0017	 & 	5.7e-129	 & 	1.2e-44	 & 	3.8e-10	 & 	100.0	 & 	3.8e-05	 & 	0.0022	 & 	3.9e-23\\ 
 -3.0	  	\hspace{0.1cm}	  	-2.5	 & 	4.1e-18	 & 	1e-09	 & 	0.0072	 & 	1.2e-99	 & 	6.9e-36	 & 	1.7e-07	 & 	95.0	 & 	0.0024	 & 	5.3	 & 	9.6e-19\\ 
 -2.5	  	\hspace{0.1cm}	  	-2.0	 & 	2.8e-15	 & 	5.8e-08	 & 	0.0082	 & 	7.9e-75	 & 	8.3e-29	 & 	7e-06	 & 	29.0	 & 	0.02	 & 	71.0	 & 	1.8e-15\\ 
 -2.0	  	\hspace{0.1cm}	  	-1.5	 & 	1.1e-11	 & 	1.5e-05	 & 	0.016	 & 	1.8e-52	 & 	1.8e-21	 & 	0.00089	 & 	6.6	 & 	0.39	 & 	93.0	 & 	1.6e-11\\ 
 -1.5	  	\hspace{0.1cm}	  	-1.0	 & 	9e-08	 & 	0.011	 & 	0.35	 & 	1.6e-33	 & 	1.1e-14	 & 	0.29	 & 	24.0	 & 	32.0	 & 	44.0	 & 	2.2e-07\\ 
 -1.0	  	\hspace{0.1cm}	  	-0.5	 & 	1.8e-05	 & 	0.22	 & 	0.47	 & 	1.4e-19	 & 	4.2e-10	 & 	2.3	 & 	7.6	 & 	89.0	 & 	0.23	 & 	5.9e-05\\ 
 -0.5	  	\hspace{0.1cm}	  	0.0	 & 	0.0021	 & 	2.4	 & 	0.28	 & 	3.2e-09	 & 	2.3e-06	 & 	6.9	 & 	0.61	 & 	90.0	 & 	3.1e-06	 & 	0.0077\\ 
 0.0	  	\hspace{0.1cm}	  	0.5	 & 	0.15	 & 	20.0	 & 	0.19	 & 	0.018	 & 	0.0019	 & 	13.0	 & 	0.054	 & 	67.0	 & 	9.2e-13	 & 	0.52\\ 
 0.5	  	\hspace{0.1cm}	  	1.0	 & 	2.7	 & 	50.0	 & 	0.074	 & 	15.0	 & 	0.083	 & 	6.9	 & 	0.0029	 & 	19.0	 & 	4.5e-21	 & 	6.8\\ 
 1.0	  	\hspace{0.1cm}	  	1.5	 & 	17.0	 & 	35.0	 & 	0.0088	 & 	22.0	 & 	0.38	 & 	0.66	 & 	3.7e-05	 & 	1.0	 & 	8.5e-32	 & 	25.0\\ 
 1.5	  	\hspace{0.1cm}	  	2.0	 & 	48.0	 & 	15.0	 & 	0.001	 & 	0.18	 & 	0.28	 & 	0.035	 & 	4.7e-07	 & 	0.036	 & 	3.2e-44	 & 	36.0\\ 
 2.0	  	\hspace{0.1cm}	  	2.5	 & 	75.0	 & 	3.2	 & 	8.1e-05	 & 	4.4e-07	 & 	0.028	 & 	0.00063	 & 	3.2e-09	 & 	0.0005	 & 	1.1e-58	 & 	22.0\\ 
 2.5	  	\hspace{0.1cm}	  	3.0	 & 	91.0	 & 	0.55	 & 	7.6e-06	 & 	3.6e-16	 & 	0.00054	 & 	7e-06	 & 	2.3e-11	 & 	5.2e-06	 & 	7.3e-75	 & 	8.5\\ 
 3.0	  	\hspace{0.1cm}	  	3.5	 & 	98.0	 & 	0.089	 & 	1e-06	 & 	9.4e-29	 & 	2.1e-06	 & 	5.5e-08	 & 	2.1e-13	 & 	4.6e-08	 & 	1.2e-92	 & 	2.3\\ 
 3.5	  	\hspace{0.1cm}	  	4.0	 & 	100.0	 & 	0.015	 & 	2e-07	 & 	8.6e-45	 & 	1.8e-09	 & 	3.4e-10	 & 	2.6e-15	 & 	3.8e-10	 & 	4.7e-112	 & 	0.46\\ 
     \hline
\end{tabular}
\end{threeparttable}
\label{probR}
\end{table*}

\begin{table*}
\caption{Probabilities distribution of variability time-scales for declining phase of the flare, 
for a range of measured decline rates (time given on logarithmic scale). These distributions have been calculated for the flux density limit of 0.1~mJy.}
\centering
\begin{threeparttable}[b]
\begin{tabular}{ccccccccccccc}
\hline
log $\tau$ (day) & AGN & SN &	XRB & GRB & Nova & RSCVn & Flare star & Algol &Mag CV & DN & TDE & Magnetar\\
\hline
-4.0	  	\hspace{0.1cm}	  	-3.5	 & 	2.4e-28	 & 	3.8e-11	 & 	0.01	 & 	3.1e-26	 & 	2.6e-59	 & 	8.2e-13	 & 	100.0	 & 	1.9e-06	 & 	2e-106	 & 	2e-19	 & 	8.7e-136	 & 	3.8e-22\\ 
 -3.5	  	\hspace{0.1cm}	  	-3.0	 & 	1.4e-24	 & 	3e-10	 & 	0.0046	 & 	1.6e-22	 & 	7.5e-51	 & 	1.7e-10	 & 	100.0	 & 	3.1e-05	 & 	1.2e-48	 & 	2.6e-16	 & 	4.3e-116	 & 	2.6e-19\\ 
 -3.0	  	\hspace{0.1cm}	  	-2.5	 & 	1.1e-20	 & 	5.8e-09	 & 	0.0051	 & 	1e-18	 & 	1.1e-42	 & 	3.8e-08	 & 	100.0	 & 	0.00072	 & 	9.9e-13	 & 	3.6e-13	 & 	1.5e-97	 & 	1.9e-16\\ 
 -2.5	  	\hspace{0.1cm}	  	-2.0	 & 	1.8e-17	 & 	1e-07	 & 	0.0086	 & 	1.3e-15	 & 	5.7e-36	 & 	3e-06	 & 	76.0	 & 	0.01	 & 	24.0	 & 	1.1e-10	 & 	1.6e-81	 & 	3.6e-14\\ 
 -2.0	  	\hspace{0.1cm}	  	-1.5	 & 	1.5e-12	 & 	2.5e-05	 & 	0.07	 & 	6.3e-11	 & 	2.8e-27	 & 	0.0029	 & 	50.0	 & 	1.1	 & 	49.0	 & 	7.4e-07	 & 	5.8e-64	 & 	1.1e-10\\ 
 -1.5	  	\hspace{0.1cm}	  	-1.0	 & 	7.7e-09	 & 	0.0023	 & 	0.66	 & 	1.8e-07	 & 	1.2e-20	 & 	0.35	 & 	62.0	 & 	37.0	 & 	4.1e-07	 & 	0.00037	 & 	1.4e-49	 & 	3.1e-08\\ 
 -1.0	  	\hspace{0.1cm}	  	-0.5	 & 	2.4e-06	 & 	0.022	 & 	0.67	 & 	3e-05	 & 	1.2e-15	 & 	2.6	 & 	4.0	 & 	93.0	 & 	2.9e-38	 & 	0.0095	 & 	1.2e-37	 & 	4.6e-07\\ 
 -0.5	  	\hspace{0.1cm}	  	0.0	 & 	0.00046	 & 	0.15	 & 	0.33	 & 	0.0025	 & 	3e-11	 & 	7.3	 & 	0.032	 & 	92.0	 & 	1.6e-92	 & 	0.1	 & 	3.9e-27	 & 	2.6e-06\\ 
 0.0	  	\hspace{0.1cm}	  	0.5	 & 	0.07	 & 	1.5	 & 	0.24	 & 	0.15	 & 	2.2e-07	 & 	15.0	 & 	0.00015	 & 	83.0	 & 	5.9e-169	 & 	0.7	 & 	4.6e-18	 & 	1e-05\\ 
 0.5	  	\hspace{0.1cm}	  	1.0	 & 	5.9	 & 	15.0	 & 	0.19	 & 	4.8	 & 	0.00035	 & 	17.0	 & 	4.2e-07	 & 	55.0	 & 	1.5e-267	 & 	2.4	 & 	1.7e-10	 & 	2e-05\\ 
 1.0	  	\hspace{0.1cm}	  	1.5	 & 	46.0	 & 	27.0	 & 	0.046	 & 	15.0	 & 	0.016	 & 	3.3	 & 	2.6e-10	 & 	7.8	 & 	0.0	 & 	1.1	 & 	1.9e-05	 & 	5.6e-06\\ 
 1.5	  	\hspace{0.1cm}	  	2.0	 & 	78.0	 & 	14.0	 & 	0.002	 & 	7.7	 & 	0.086	 & 	0.061	 & 	7.9e-15	 & 	0.12	 & 	0.0	 & 	0.057	 & 	0.042	 & 	1.6e-07\\ 
 2.0	  	\hspace{0.1cm}	  	2.5	 & 	86.0	 & 	7.0	 & 	8e-05	 & 	2.3	 & 	0.13	 & 	0.0005	 & 	8.1e-20	 & 	0.0011	 & 	0.0	 & 	0.0012	 & 	4.3	 & 	1.8e-09\\ 
 2.5	  	\hspace{0.1cm}	  	3.0	 & 	68.0	 & 	4.3	 & 	3.9e-06	 & 	0.41	 & 	0.062	 & 	2.5e-06	 & 	4.4e-25	 & 	7.8e-06	 & 	0.0	 & 	1.5e-05	 & 	27.0	 & 	1.2e-11\\ 
 3.0	  	\hspace{0.1cm}	  	3.5	 & 	72.0	 & 	6.4	 & 	3.9e-07	 & 	0.081	 & 	0.014	 & 	1.1e-08	 & 	1.6e-30	 & 	6.1e-08	 & 	0.0	 & 	1.5e-07	 & 	22.0	 & 	6.3e-14\\ 
 3.5	  	\hspace{0.1cm}	  	4.0	 & 	81.0	 & 	17.0	 & 	8.6e-08	 & 	0.017	 & 	0.0016	 & 	5.9e-11	 & 	7.2e-36	 & 	7.6e-10	 & 	0.0	 & 	1.7e-09	 & 	1.8	 & 	3.8e-16\\ 
     \hline
\end{tabular}
\end{threeparttable}
\label{probD}
\end{table*}

\section{Additional figures}
\label{additional_figures}

\begin{figure*}
\includegraphics[width=0.49\textwidth]{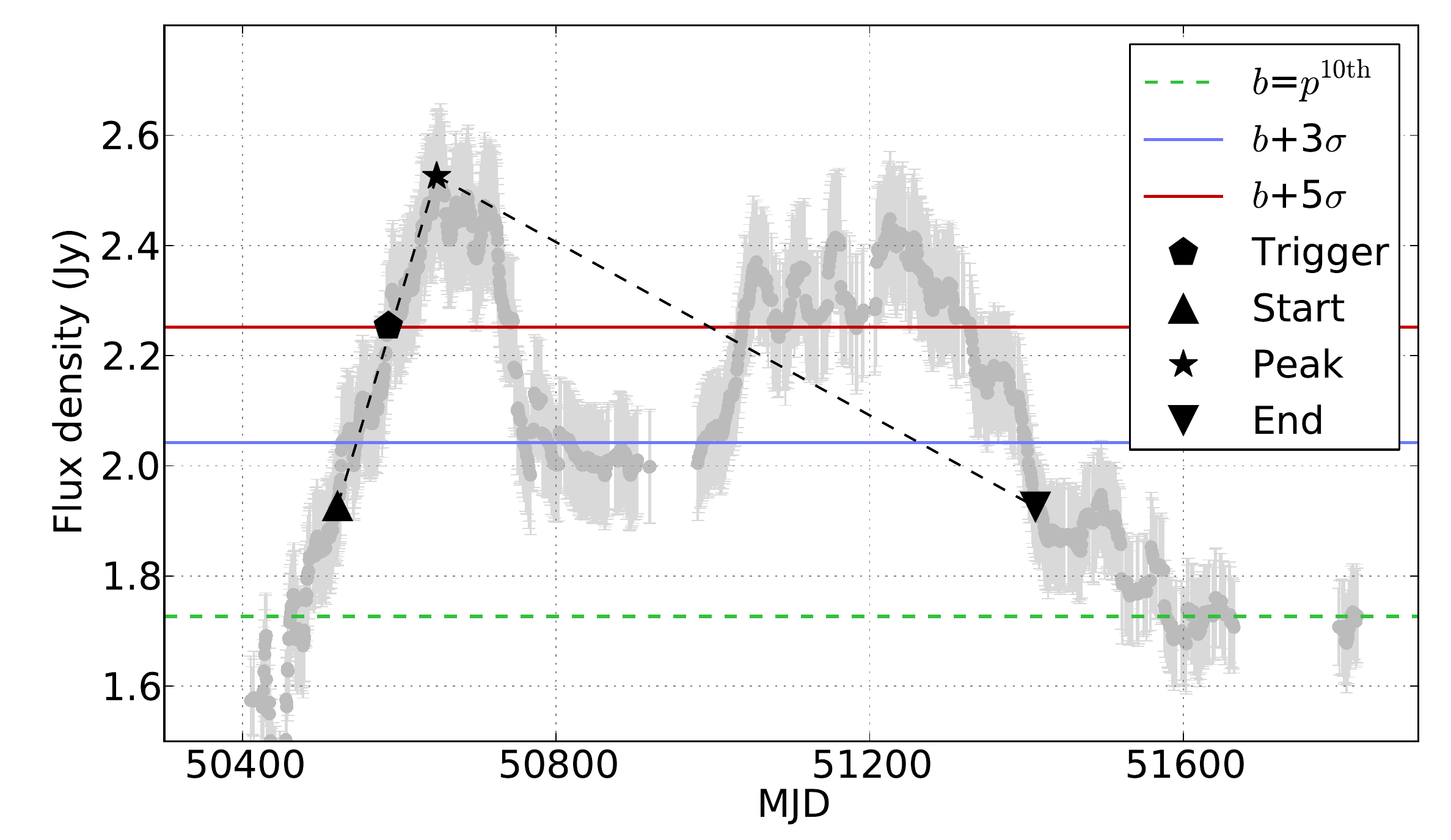}
\includegraphics[width=0.493\textwidth]{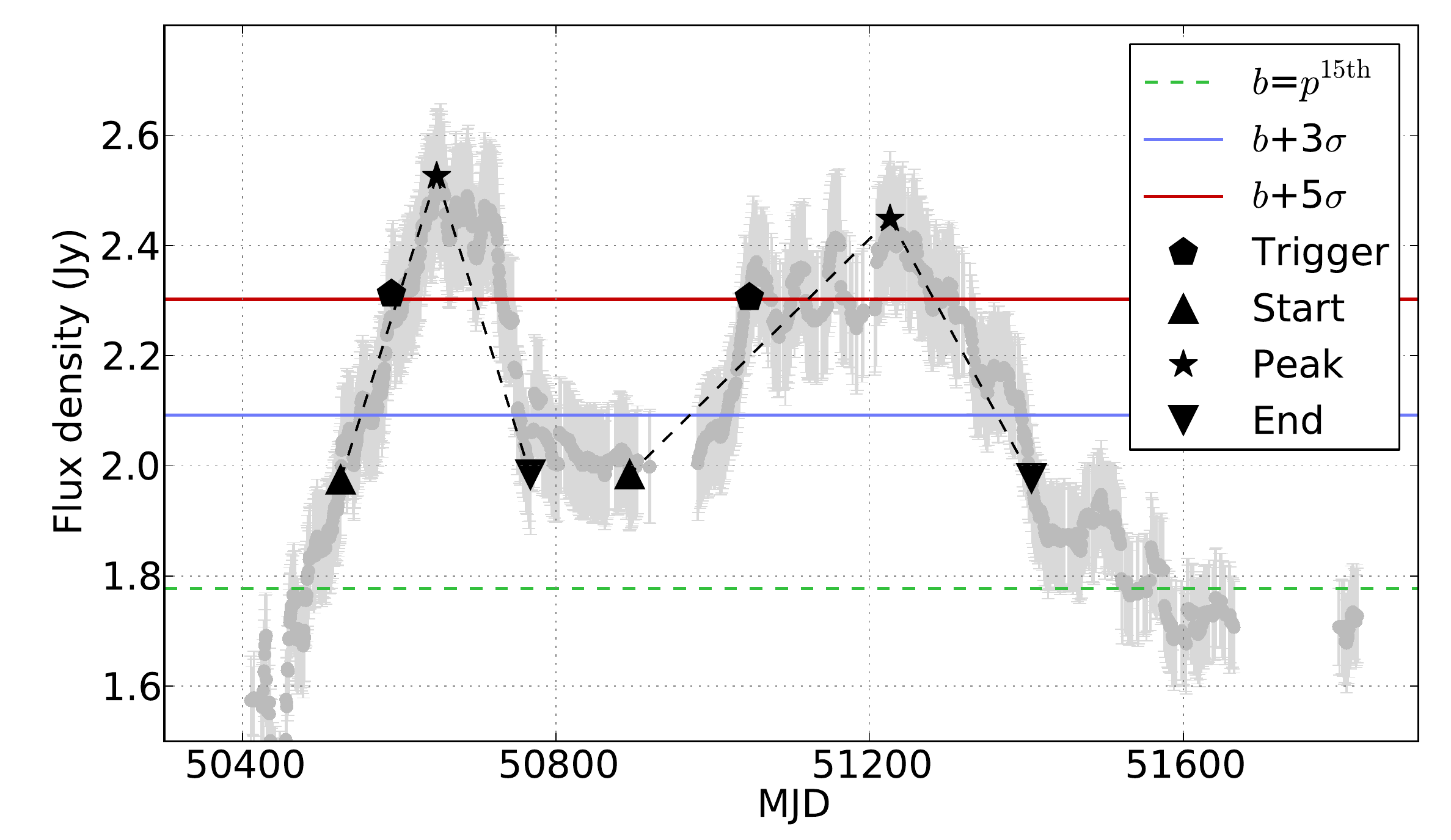}
\includegraphics[width=0.49\textwidth]{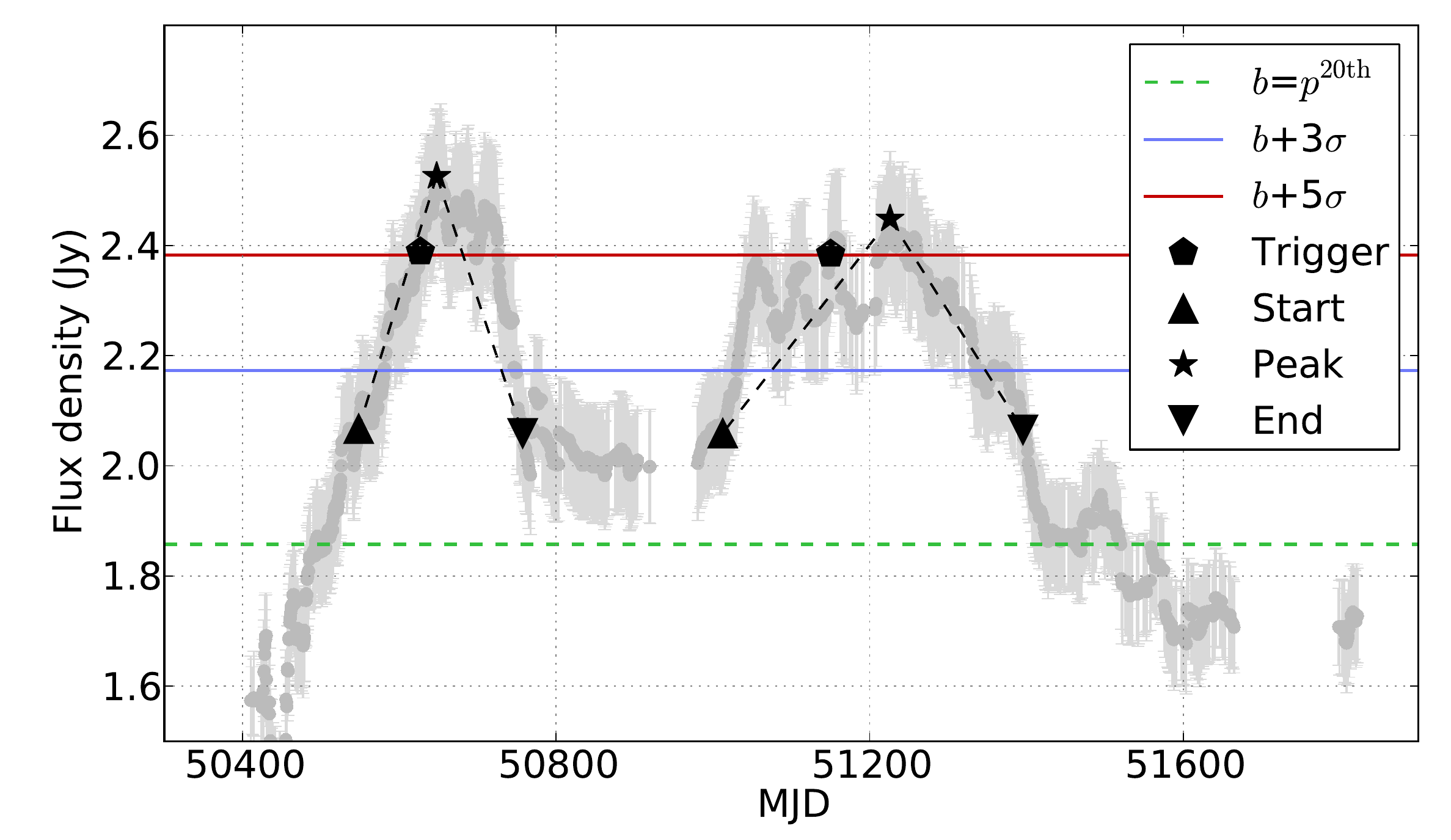}
\includegraphics[width=0.493\textwidth]{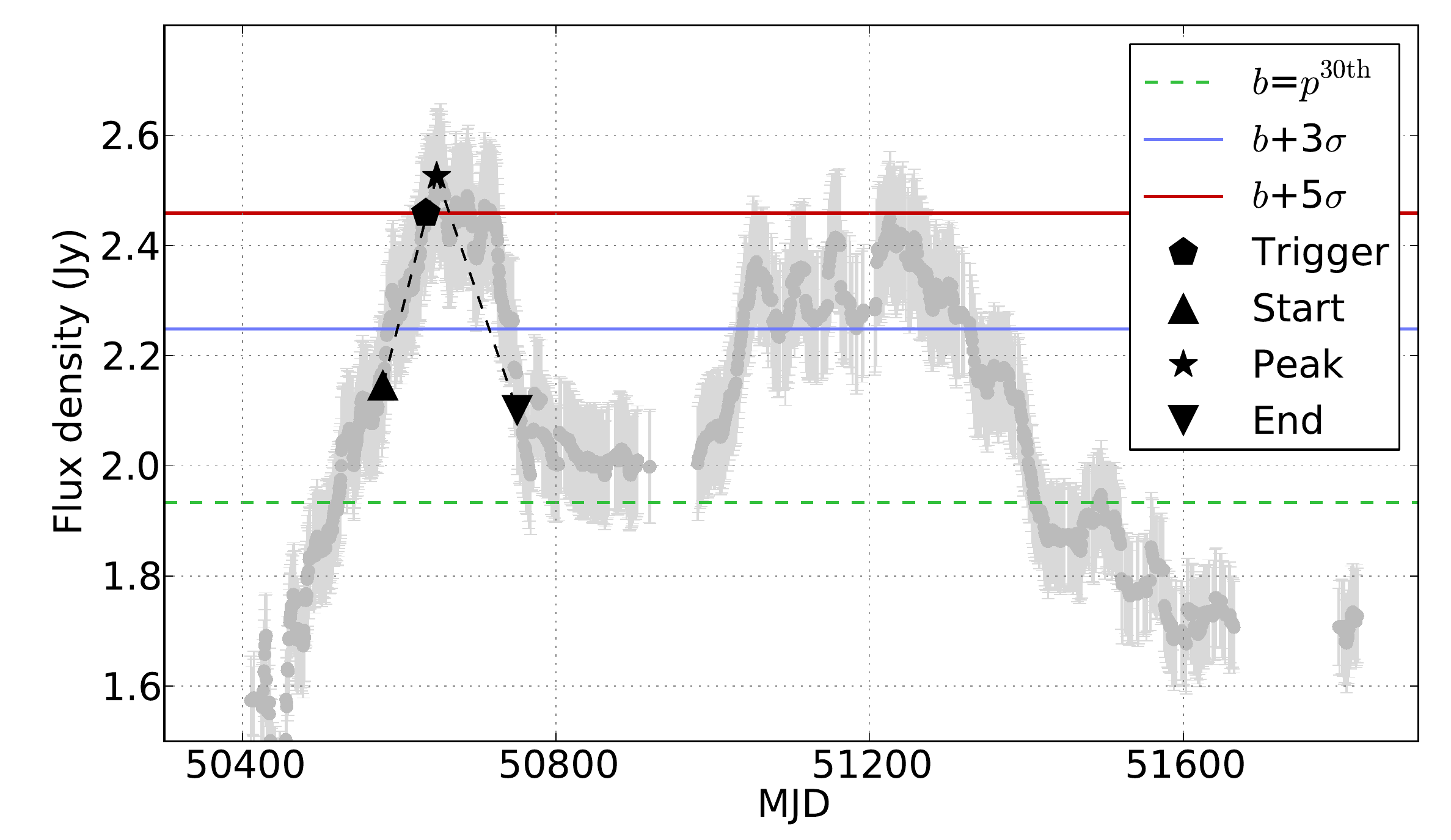}
\caption{Selection of flares from a light-curve of blazar 0336-019 for four progressively higher estimations of the background level: 10th, 15th, 20th and 30th percentile of the flux density measurements. In each case, dashed lines illustrate the respective parts of flares for which the measurement of rise/decline time-scale is made. {\em Upper left}: two flares mistakenly interpreted as a single outburst, with overestimated decline rate as a result; {Upper right and lower left}: individual flares detected separately; {Lower right}: fainter of the flares below the detection threshold. }
\label{missed-flare-example}
\end{figure*}

\begin{figure*}
\includegraphics[width=0.49\textwidth]{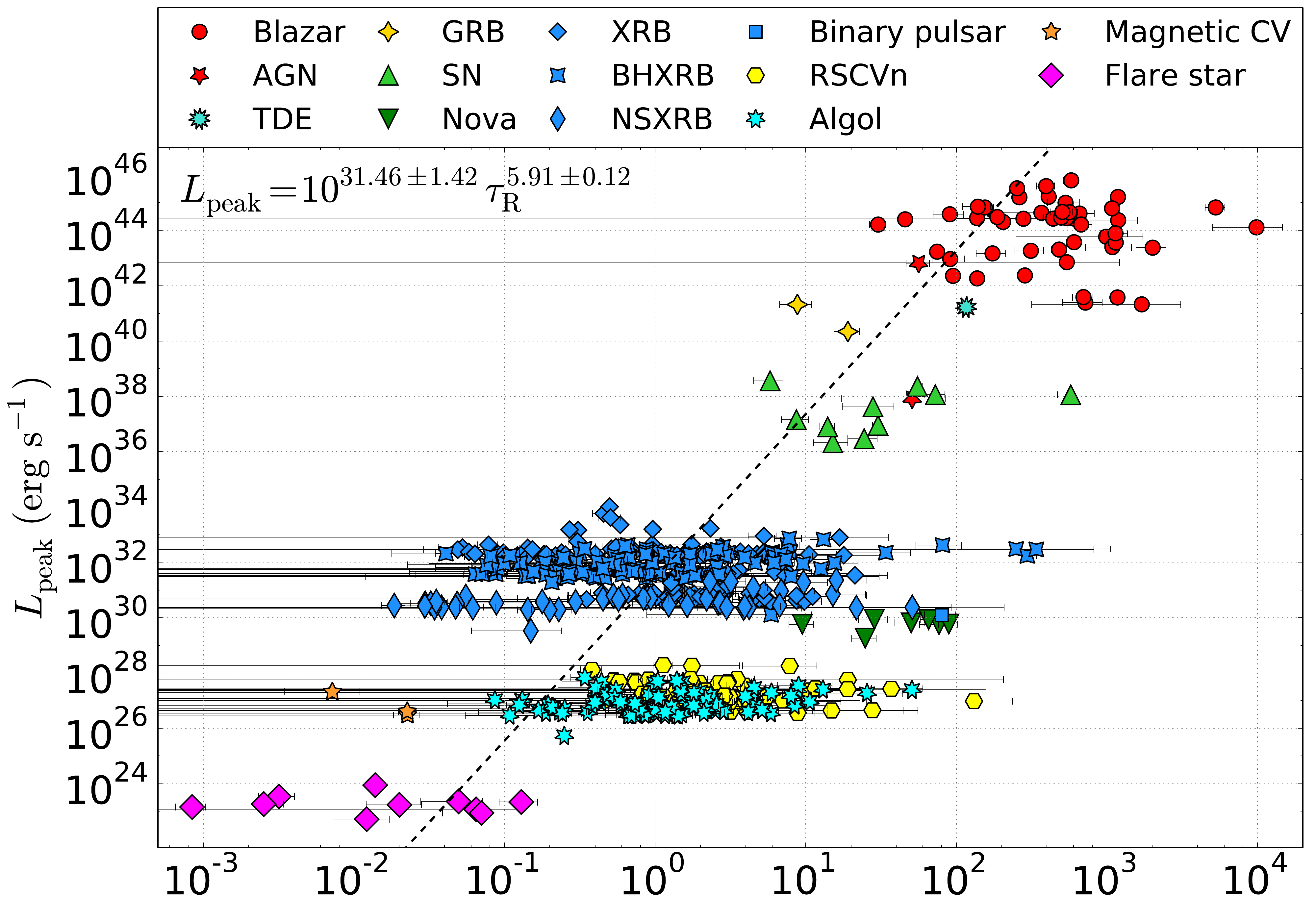}
\includegraphics[width=0.493\textwidth]{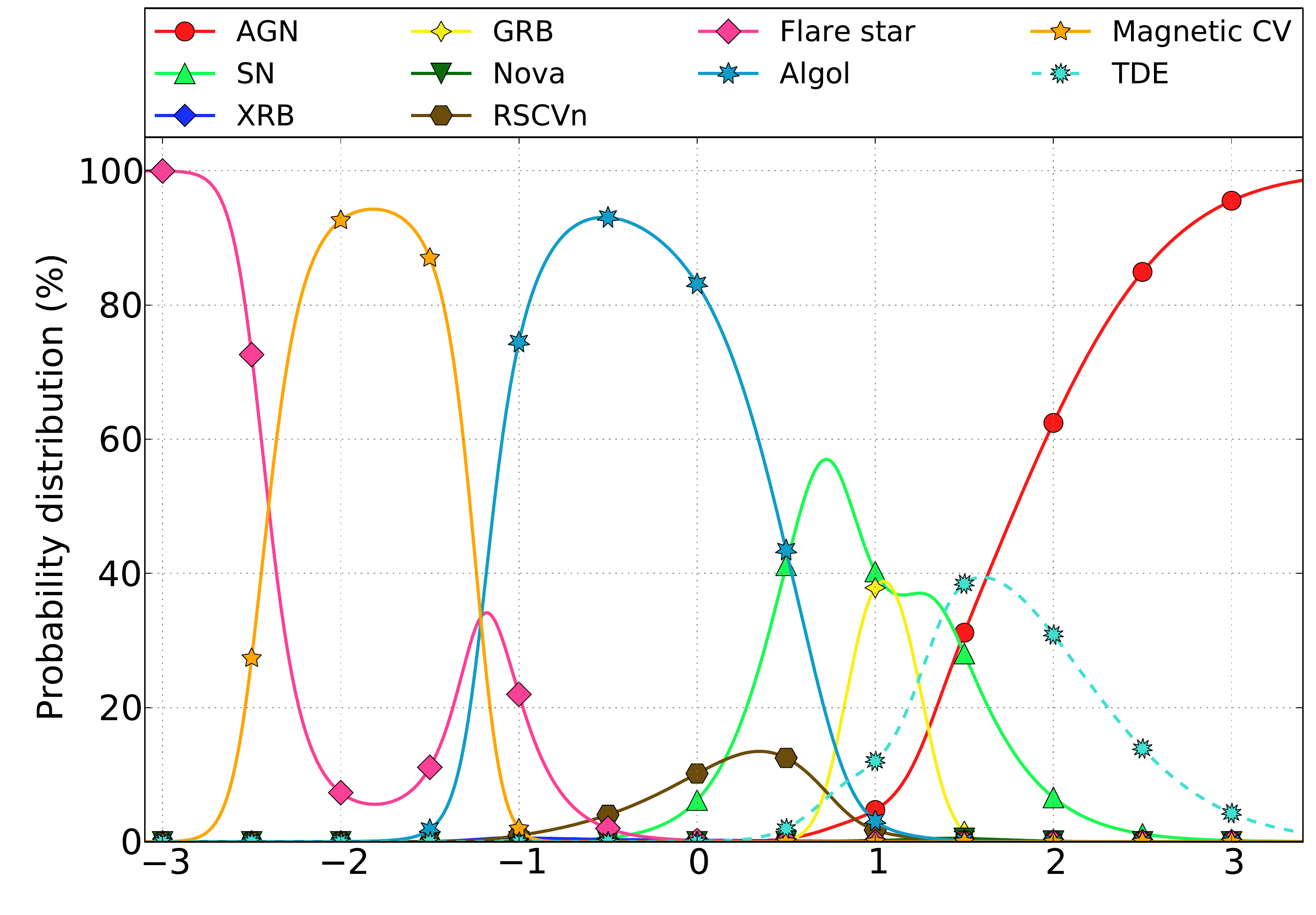}
\includegraphics[width=0.49\textwidth]{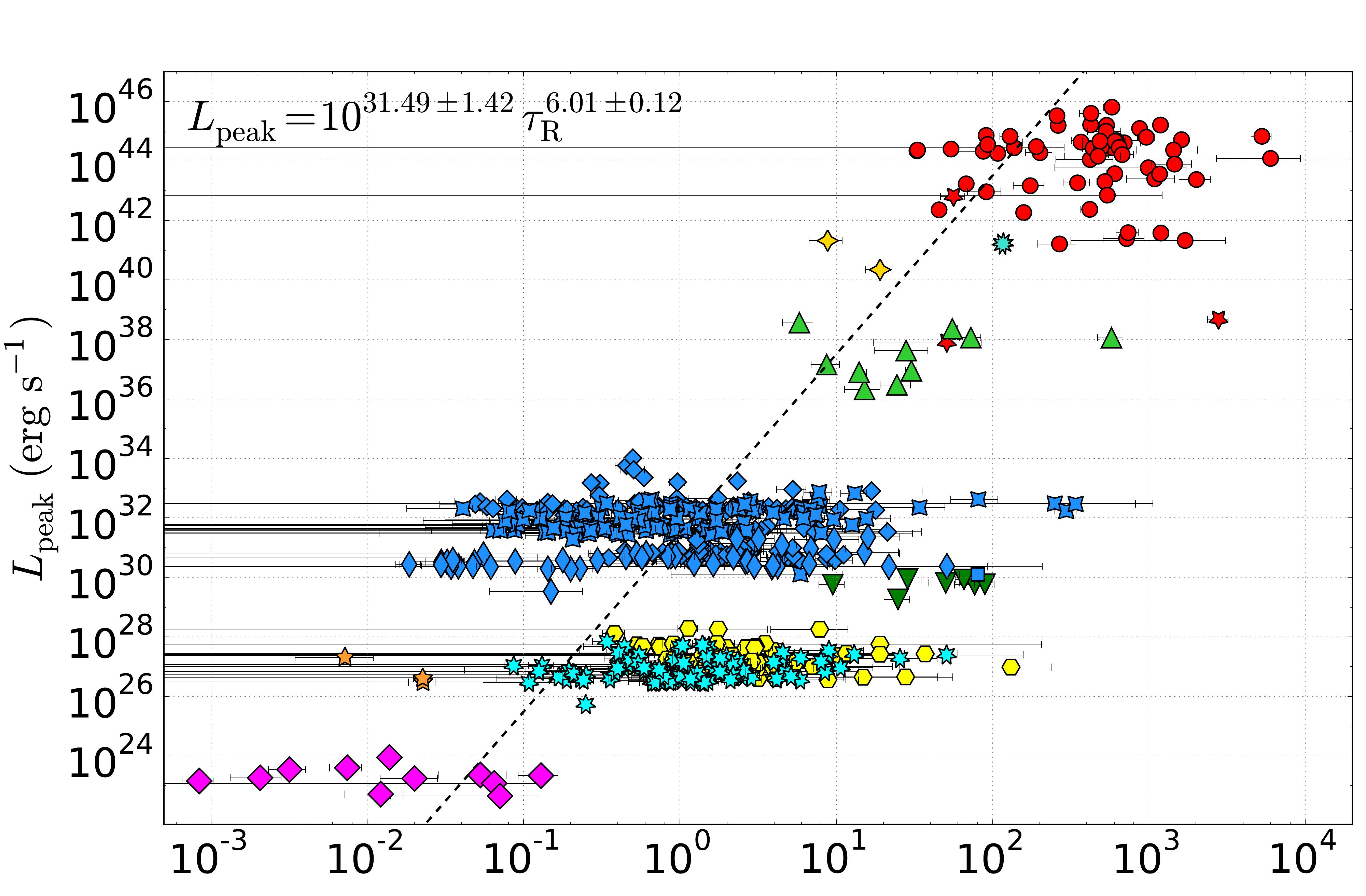}
\includegraphics[width=0.493\textwidth]{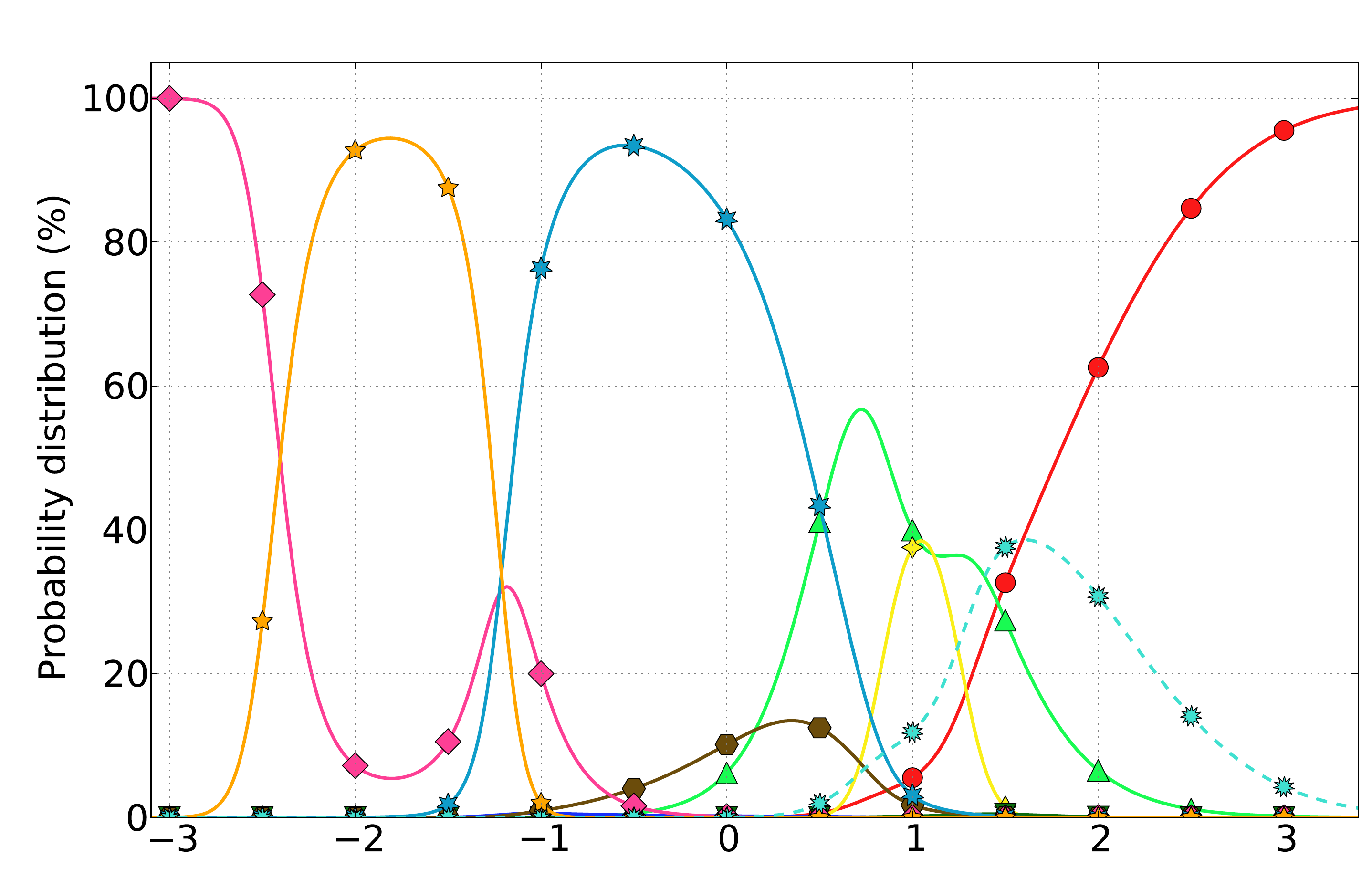}
\includegraphics[width=0.49\textwidth]{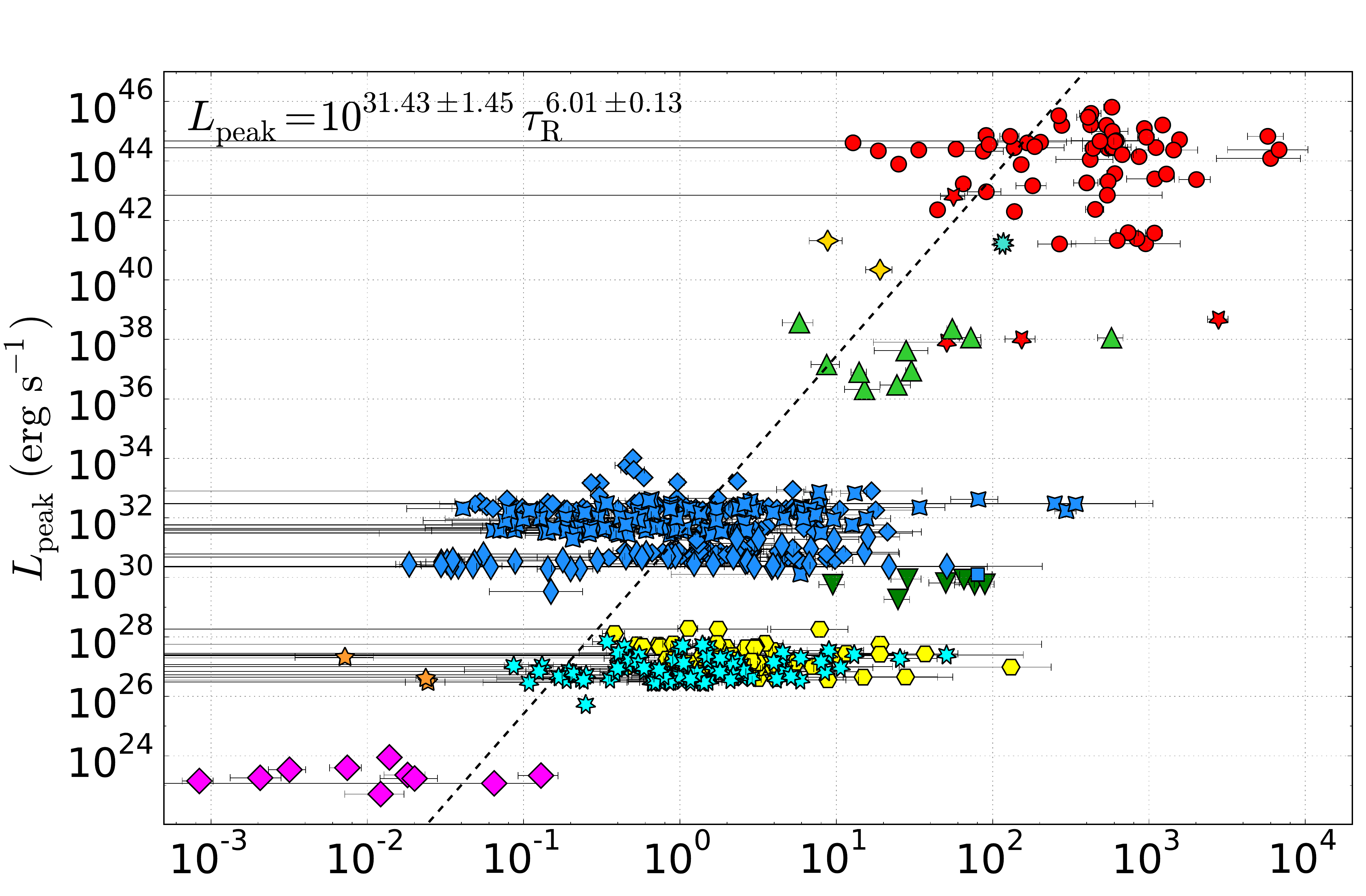}
\includegraphics[width=0.493\textwidth]{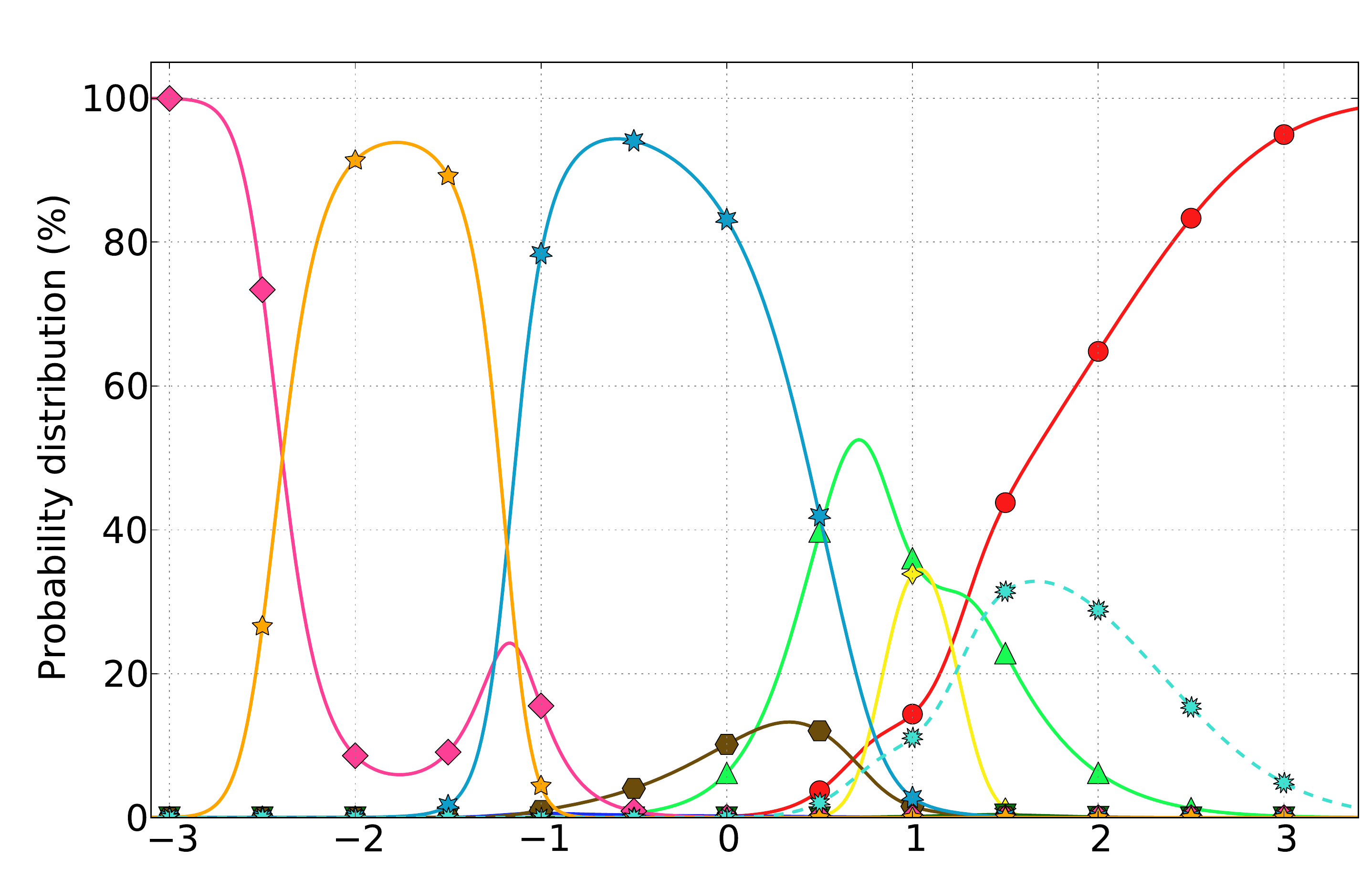}
\includegraphics[width=0.49\textwidth]{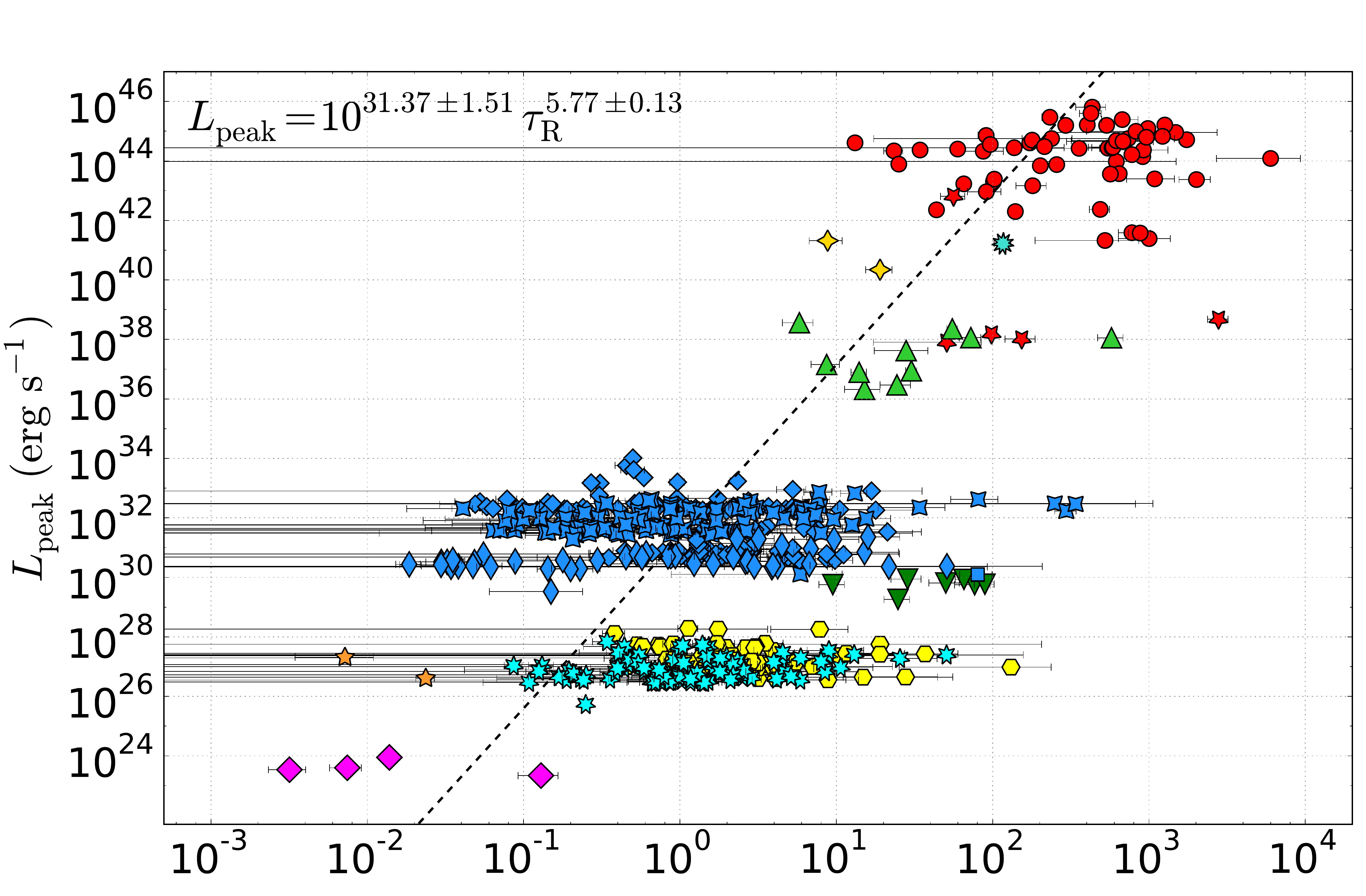}
\includegraphics[width=0.493\textwidth]{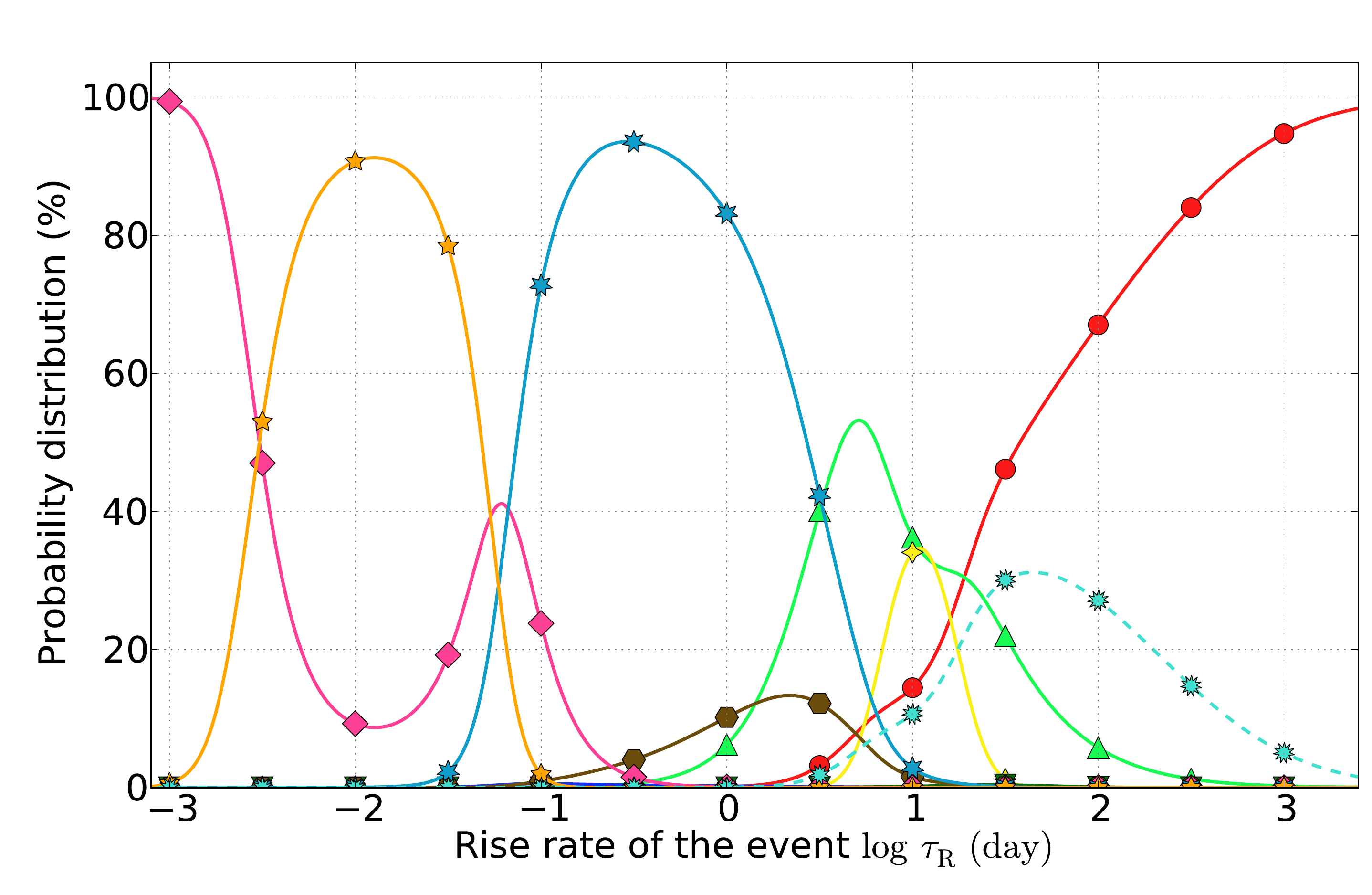}
\caption{Variation of Figures \ref{LT} and lower panel of Figure \ref{RP} obtained for a number of different background estimation parameters, as discussed in Sections \ref{datalits}, \ref{datagbis}: from top to bottom, background estimated as 10, 15, 20, 30th percentile of the flux density values of the light-curve.}
\label{background-est-var}
\end{figure*}

\begin{figure*}
\includegraphics[width=\textwidth]{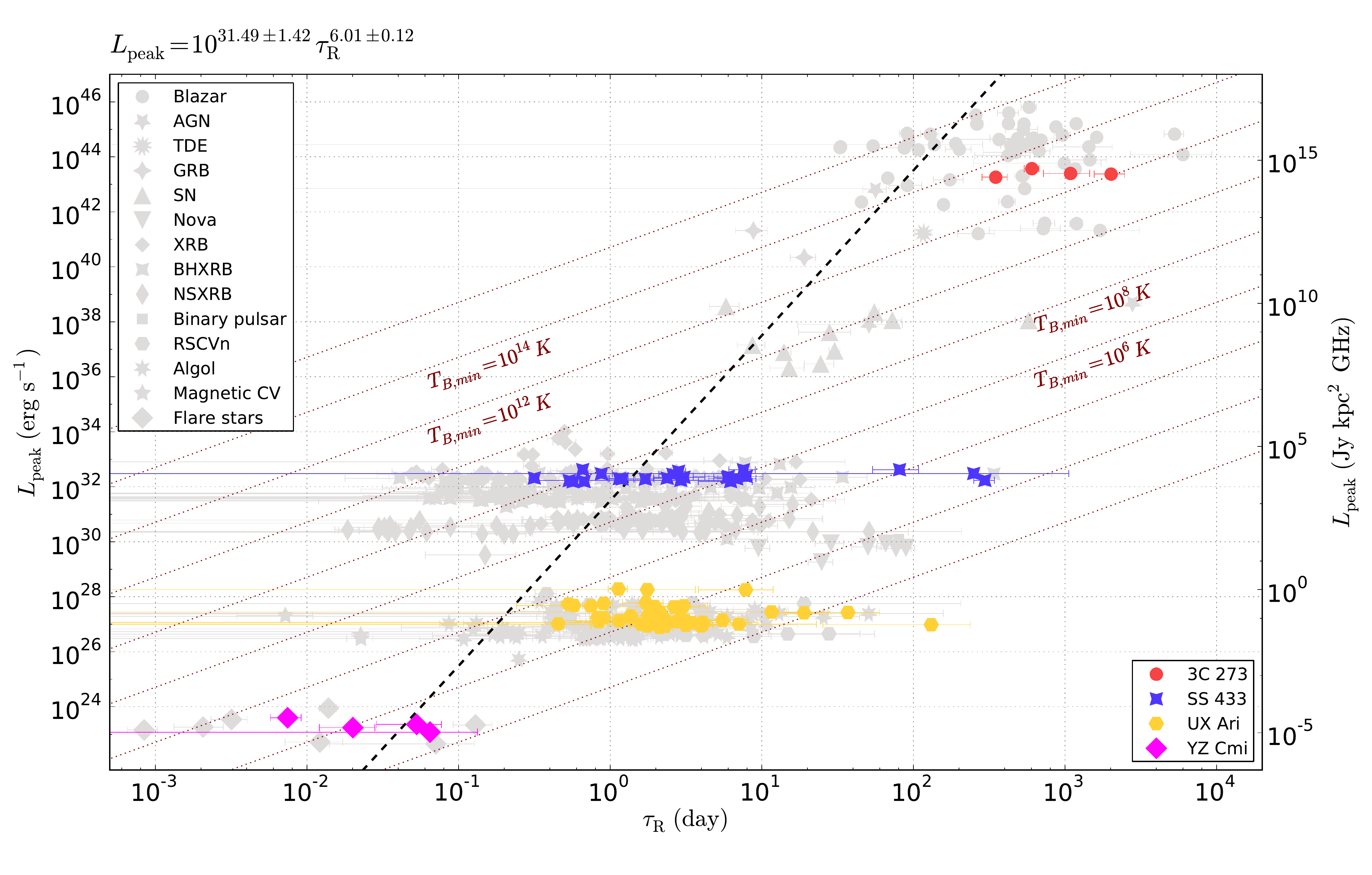}
\caption{Variation of Figure \ref{LT} showing the scatter of measured time-scale of rise rates for individual sources.}
\label{single-source-scatter}
\end{figure*}


\begin{thebibliography}{}

\bibitem[\protect\citeauthoryear{Alexander et al.}{2016}]{14li} Alexander K.~D., Berger E., Guillochon J., Zauderer B.~A., Williams P.~K.~G., 2016, ApJ, 819, L25 

\bibitem[\protect\citeauthoryear{Astropy Collaboration et 
al.}{2013}]{astropy} Astropy Collaboration, et al., 2013, A\&A, 558, A33 

\bibitem[\protect\citeauthoryear{Bower et al.}{2007}]{Bower2007} 
Bower G.~C., Saul D., Bloom J.~S., Bolatto A., Filippenko A.~V., Foley 
R.~J., Perley D., 2007, ApJ, 666, 346 

\bibitem[\protect\citeauthoryear{de Bruyn 
\& Macquart}{2015}]{1819+3845D} de Bruyn A.~G., Macquart J.-P., 2015, A\&A, 574, AA125 

\bibitem[\protect\citeauthoryear{Burlon et al.}{2015}]{Burlon2015} Burlon D., Ghirlanda G., van der Horst A., Murphy T., Wijers R.~A.~M.~J., Gaensler B., Ghisellini G., Prandoni I., 2015, aska.conf, 52 

\bibitem[\protect\citeauthoryear{Coppejans et 
al.}{2015}]{coppejans2015} Coppejans D.~L., K{\"o}rding E.~G., 
Miller-Jones J.~C.~A., Rupen M.~P., Knigge C., Sivakoff G.~R., Groot P.~J., 
2015, MNRAS, 451, 3801 

\bibitem[\protect\citeauthoryear{Dennett-Thorpe 
\& de Bruyn}{2003}]{1819+3845} Dennett-Thorpe J., de Bruyn A.~G., 2003, A\&A, 404, 113 

\bibitem[\protect\citeauthoryear{Djorgovski et 
al.}{2012}]{djorgovski2012} Djorgovski S.~G., Mahabal A.~A., Donalek 
C., Graham M.~J., Drake A.~J., Moghaddam B., Turmon M., 2012, arXiv, 
arXiv:1209.1681 

\bibitem[\protect\citeauthoryear{Duerbeck}{1984}]{Duerbeck1984} Duerbeck H.~W., 1984, Ap\&SS, 99, 363 

\bibitem[\protect\citeauthoryear{Favata, Micela, \& Sciortino}{1995}]{Favata1995} Favata F., Micela G., Sciortino S., 1995, A\&A, 298, 482 

\bibitem[\protect\citeauthoryear{Fender et al.}{2015}]{Fender2015} 
Fender R., Stewart A., Macquart J.-P., Donnarumma I., Murphy T., Deller A., 
Paragi Z., Chatterjee S., 2015, arXiv, arXiv:1507.00729 

\bibitem[\protect\citeauthoryear{Fiedler et 
al.}{1987}]{ESE1} Fiedler R.~L., Dennison B., Johnston 
K.~J., Hewish A., 1987, Natur, 326, 675 


\bibitem[\protect\citeauthoryear{Findeisen, Cody, 
\& Hillenbrand}{2015}]{findeisen2015} Findeisen K., Cody A.~M., Hillenbrand L., 2015, ApJ, 798, 89 

\bibitem[\protect\citeauthoryear{Frail et al.}{2012}]{Frail2012} 
Frail D.~A., Kulkarni S.~R., Ofek E.~O., Bower G.~C., Nakar E., 2012, ApJ, 
747, 70 

\bibitem[\protect\citeauthoryear{Gab{\'a}nyi et 
al.}{2007}]{J1128+5925} Gab{\'a}nyi K.~{\'E}., et al., 2007, A\&A, 470, 83 

\bibitem[\protect\citeauthoryear{Gallo, Fender, 
\& Pooley}{2003}]{xrbrate} Gallo E., Fender R.~P., Pooley G.~G., 2003, MNRAS, 344, 60 

\bibitem[\protect\citeauthoryear{Hilton et al.}{2010}]{Hilton2010} Hilton E.~J., West A.~A., Hawley S.~L., Kowalski A.~F., 2010, AJ, 140, 1402 

\bibitem[\protect\citeauthoryear{Huppenkothen et 
al.}{2015}]{Huppenkothen2015} Huppenkothen D., et al., 2015, ApJ, 810, 
66 

\bibitem[\protect\citeauthoryear{Jauncey et 
al.}{2003}]{1257-326} Jauncey D.~L., Bignall H.~E., Lovell 
J.~E.~J., Kedziora-Chudczer L., Tzioumis A.~K., Macquart J.-P., Rickett 
B.~J., 2003, ASPC, 300, 199 

\bibitem[\protect\citeauthoryear{Koay et 
al.}{2011}]{1328+6221} Koay J.~Y., Bignall H.~E., Macquart J.-P., Jauncey D.~L., Rickett B.~J., Lovell J.~E.~J., 2011, A\&A, 534, LL1 

\bibitem[\protect\citeauthoryear{Koay et al.}{2011}]{KoaySCT} 
Koay J.~Y., et al., 2011, AJ, 142, 108 

\bibitem[\protect\citeauthoryear{K{\"o}rding et 
al.}{2011}]{kording2011} K{\"o}rding E.~G., Knigge C., Tzioumis T., 
Fender R., 2011, MNRAS, 418, L129 

\bibitem[\protect\citeauthoryear{K{\"u}gler, Gianniotis, 
\& Polsterer}{2015}]{kugler} K{\"u}gler S.~D., Gianniotis N., Polsterer K.~L., 2015, MNRAS, 451, 3385 

\bibitem[\protect\citeauthoryear{Lo et al.}{2014}]{Lo2014} Lo 
K.~K., Murphy T., Rebbapragada U., Wagstaff K., 2014, arXiv, 
arXiv:1402.7180 

\bibitem[\protect\citeauthoryear{Lo et al.}{2014}]{Lo} Lo 
K.~K., Farrell S., Murphy T., Gaensler B.~M., 2014, ApJ, 786, 20 

\bibitem[\protect\citeauthoryear{Lovell et al.}{2003}]{Lovell2003} Lovell J.~E.~J., Jauncey D.~L., Bignall H.~E., Kedziora-Chudczer L., Macquart J.-P., Rickett B.~J., Tzioumis A.~K., 2003, AJ, 126, 1699

\bibitem[\protect\citeauthoryear{Metzger, Williams, \& Berger}{2015}]{Metzger2015} Metzger B.~D., Williams P.~K.~G., Berger E., 2015, ApJ, 806, 224 

\bibitem[\protect\citeauthoryear{O'Brien et 
al.}{2006}]{RSOph} O'Brien T., et al., 2006, evn..conf,  

\bibitem[\protect\citeauthoryear{Ofek}{2007}]{Ofek2007} Ofek E.~O., 2007, ApJ, 659, 339 

\bibitem[\protect\citeauthoryear{Olausen \& Kaspi}{2014}]{Olausen2014} Olausen S.~A., Kaspi V.~M., 2014, ApJS, 212, 6 

\bibitem[\protect\citeauthoryear{Osten}{2008}]{Osten2008} Osten R.~A., 2008, arXiv, arXiv:0801.2573 

\bibitem[\protect\citeauthoryear{Ottmann 
\& Schmitt}{1992}]{Ottmann1992} Ottmann R., Schmitt J.~H.~M.~M., 1992, A\&A, 256, 421 


\bibitem[\protect\citeauthoryear{Perez-Torres et al.}{2015}]{Perez-Torres2015} Perez-Torres M., et al., 2015, aska.conf, 60 


\bibitem[\protect\citeauthoryear{Pichara 
\& Protopapas}{2013}]{pichara} Pichara K., Protopapas P., 2013, ApJ, 777, 83 


\bibitem[\protect\citeauthoryear{Pietka, Fender, 
\& Keane}{2015}]{paper1} Pietka M., Fender R.~P., Keane E.~F., 2015, MNRAS, 446, 3687 

\bibitem[\protect\citeauthoryear{Pitkin et al.}{2014}]{pitkin2014} 
Pitkin M., Williams D., Fletcher L., Grant S.~D.~T., 2014, MNRAS, 445, 2268 

\bibitem[\protect\citeauthoryear{Pretorius 
\& Knigge}{2012}]{PK2012} Pretorius M.~L., Knigge C., 2012, MNRAS, 419, 1442 

\bibitem[\protect\citeauthoryear{Pretorius, Knigge, 
\& Schwope}{2013}]{Pretorius2013} Pretorius M.~L., Knigge C., Schwope A.~D., 2013, MNRAS, 432, 570 

\bibitem[\protect\citeauthoryear{Pretorius 
\& Mukai}{2014}]{ipspacedens} Pretorius M.~L., Mukai K., 2014, MNRAS, 442, 2580 

\bibitem[\protect\citeauthoryear{Ramsay et al.}{2004}]{polarsdc} 
Ramsay G., Cropper M., Wu K., Mason K.~O., C{\'o}rdova F.~A., Priedhorsky 
W., 2004, MNRAS, 350, 1373 

\bibitem[\protect\citeauthoryear{Rau et al.}{2009}]{Rau2009} Rau A., et al., 2009, PASP, 121, 1334 

\bibitem[\protect\citeauthoryear{Rebbapragada et 
al.}{2012}]{rebbapragada} Rebbapragada U., Lo K., Wagstaff K.~L., 
Reed C., Murphy T., Thompson D.~R., 2012, IAUS, 285, 397 

\bibitem[\protect\citeauthoryear{Reid, Cruz, \& Allen}{2007}]{Reid2007} Reid I.~N., Cruz K.~L., Allen P.~R., 2007, AJ, 133, 2825 

\bibitem[\protect\citeauthoryear{Richards et 
al.}{2011}]{richards} Richards J.~W., et al., 2011, ApJ, 733, 10 

\bibitem[\protect\citeauthoryear{Rickett, Lazio, 
\& Ghigo}{2006}]{sctexp} Rickett B.~J., Lazio T.~J.~W., Ghigo F.~D., 2006, ApJS, 165, 439 

\bibitem[\protect\citeauthoryear{Rickett}{2007}]{0405-385} 
Rickett B.~J., 2007, RMxAC, 27, 129 

\bibitem[\protect\citeauthoryear{Rickett et 
al.}{1995}]{0917+624} Rickett B.~J., Quirrenbach A., Wegner R., Krichbaum T.~P., Witzel A., 1995, A\&A, 293, 479 

\bibitem[\protect\citeauthoryear{Roy et al.}{2012}]{Roy2012} 
Roy N., et al., 2012, BASI, 40, 293 

\bibitem[\protect\citeauthoryear{Saglia et al.}{2012}]{saglia2012} 
Saglia R.~P., et al., 2012, ApJ, 746, 128 

\bibitem[\protect\citeauthoryear{Scargle}{1998}]{scargle1998} 
Scargle J.~D., 1998, ApJ, 504, 405 

\bibitem[\protect\citeauthoryear{Senkbeil et 
al.}{2008}]{ESE2} Senkbeil C.~E., Ellingsen S.~P., Lovell 
J.~E.~J., Macquart J.-P., Cim{\`o} G., Jauncey D.~L., 2008, ApJ, 672, L95 

\bibitem[\protect\citeauthoryear{Servillat et 
al.}{2011}]{Servillat2011} Servillat M., Webb N.~A., Lewis F., Knigge 
C., van den Berg M., Dieball A., Grindlay J., 2011, ApJ, 733, 106 



\bibitem[\protect\citeauthoryear{Stewart et al.}{2016}]{Stewart2016} Stewart A.~J., et al., 2016, MNRAS, 456, 2321 



\bibitem[\protect\citeauthoryear{Swinbank et 
al.}{2015}]{trap} Swinbank J.~D., et al., 2015, A\&C, 11, 25 

\bibitem[\protect\citeauthoryear{Thyagarajan et 
al.}{2011}]{FIRST} Thyagarajan N., Helfand D.~J., White 
R.~L., Becker R.~H., 2011, ApJ, 742, 49 

\bibitem[\protect\citeauthoryear{Tinney, Reid, 
\& Mould}{1993}]{vmax} Tinney C.~G., Reid I.~N., Mould J.~R., 1993, ApJ, 414, 254 

\bibitem[\protect\citeauthoryear{Turner et al.}{2012}]{1144-379} 
Turner R.~J., Ellingsen S.~P., Shabala S.~S., Blanchard J., Lovell 
J.~E.~J., McCallum J.~N., Cim{\`o} G., 2012, ApJ, 754, L19 

\bibitem[\protect\citeauthoryear{Williams et 
al.}{2013}]{Williams2013} Williams P.~K.~G., Bower G.~C., Croft S., 
Keating G.~K., Law C.~J., Wright M.~C.~H., 2013, ApJ, 762, 85 

\bibitem[\protect\citeauthoryear{van der Laan}{1966}]{vdl} 
van der Laan H., 1966, Natur, 211, 1131 

\bibitem[\protect\citeauthoryear{Varughese et 
al.}{2015}]{varughese} Varughese M.~M., von Sachs R., Stephanou 
M., Bassett B.~A., 2015, MNRAS, 453, 2848 

\bibitem[\protect\citeauthoryear{Volvach et 
al.}{2010}]{3C} Volvach A.~E., Ryabov M.~I., Volvach 
L.~N., Suharev A.~I., Aller H.~D., Aller M.~F., 2010, AIPC, 1206, 360 

\bibitem[\protect\citeauthoryear{Zauderer et 
al.}{2011}]{J1644+57} Zauderer B.~A., et al., 2011, Natur, 476, 
425 


\end{thebibliography}
\end{document}